\documentclass[prd,aps,nofootinbib]{revtex4}
\usepackage{amssymb,amsmath,epsfig,bm}
\usepackage[bookmarks]{hyperref}
\usepackage[usenames,dvipsnames]{color}

\newcommand{\Real}{\mathbb{R}}
\newcommand{\Complex}{\mathbb{C}}

\newcommand{\divrg}{\mbox{div}}
\newcommand{\Hz}{\mbox{\em \r{H}}}
\newcommand{\Hbz}{\mbox{\bf\em \r{H}}}

\newcommand{\proof}{\noindent {\bf Proof. }}
\newcommand{\proofof}[1]{\noindent {\bf Proof of #1. }}
\newcommand{\qed}{\hfill \fbox{} \vspace{.3cm}}

\newtheorem{lemma}{Lemma}
\newtheorem{proposition}{Proposition}

\begin{document}
%%%%%%%%%%%%%%%%%%%%%%%%%%%%%%%

\title{Linear perturbations of self-gravitating spherically symmetric configurations}
\author{Eliana Chaverra, N\'estor Ortiz, Olivier Sarbach}
\affiliation{Instituto de F\'\i sica y Matem\'aticas,
Universidad Michoacana de San Nicol\'as de Hidalgo,\\
Edificio C-3, Ciudad Universitaria, 58040 Morelia, Michoac\'an, M\'exico.}

\begin{abstract}
We present a new covariant, gauge-invariant formalism describing linear 
metric perturbation fields on any spherically symmetric background in general relativity. The advantage of this formalism relies in the fact that it does not require a decomposition of the perturbations into spherical tensor harmonics. Furthermore, it does not assume the background to be vacuum, nor does it require its staticity. In the particular case of vacuum perturbations, we derive two master equations describing the propagation of arbitrary linear gravitational waves on a Schwarzschild black hole. When decomposed into spherical harmonics, they reduce to covariant generalizations of the well-known Regge-Wheeler and Zerilli equations. Next, we discuss the general case where the metric perturbations are coupled to matter fields and derive a new constrained wave system describing the propagation of three gauge-invariant scalars from which the complete metric perturbations can be reconstructed. We apply our formalism to the Einstein-Euler system, dividing the fluid perturbations into two parts. The first part, which decouples from the metric perturbations, obeys simple advection equations along the background flow and describes the propagation of the entropy and the vorticity. The second part describes a perturbed potential flow, and together with the metric perturbations it forms a closed wave system.
\end{abstract}

\date{\today}

\pacs{04.20.-q,04.25.-g,04.40.-b}
% 04.20.-q: Classical GR
% 04.20.Ex: Initial value problem, existence and uniqueness of solutions
% 04.25.-g: Approximation methods; equations of motion
% 04.25.D-: Numerical relativity
% 04.25.Nx: Post-Newtonian approximation; perturbation theory; related approximations
% 04.40.-b: Self-gravitating systems, continuous media and classical fields in curved spacetime
% 04.70.-s: Physics of black holes
% 97.60.Lf: Astronomy: Late stage of star evolution: black holes

\maketitle

%%%%%%%%%%%%%%%%%%%%%%%%%%%%%%%%%%%%%%%%%%%
\section{Introduction}
%%%%%%%%%%%%%%%%%%%%%%%%%%%%%%%%%%%%%%%%%%%

The purpose of this work is to develop a gauge-invariant perturbation formalism describing the propagation of linearized gravitational and matter fields on an arbitrary spherically symmetric background configuration. Examples of such configurations include nonrotating black holes and stars, spherical matter distributions undergoing complete gravitational collapse, or expanding shells of matter. Therefore, it is clear that a perturbation formalism for such backgrounds possesses a wide range of applications, covering the stability analysis of compact objects, the description of quasi-normal oscillations of these objects, the computation of linearized gravitational waves produced by a small star moving in the field of a nonrotating black hole or by another black hole in the close limit approximation, the stability analysis of Cauchy horizons in gravitational collapse models, and the linear evolution of structures in cosmological models.\footnote{See Ref.~\cite{aNlR05} for a review on perturbation theory for Schwarzschild black holes and references to some of the applications we mention.}

Linear gravitational fluctuations on a Schwarzschild black hole were first studied by Regge and Wheeler~\cite{tRjW57} and Zerilli~\cite{fZ70}, who decomposed the perturbations into spherical tensor harmonics, analyzed their behavior with respect to infinitesimal coordinate transformations, and by fixing the gauge, obtained a family of master equations describing odd- and even-parity perturbations with a given angular momentum number $\ell\geq 2$. These master equations, known as the Regge-Wheeler and the Zerilli equations, respectively, describe the propagation of the true dynamical degrees of freedom of the linear theory, completely eliminating gauge and constraint modes. Therefore, they are ideally suited for understanding the behavior of linear perturbations of the Schwarzschild black hole. From the fact that these equations can be written as a $1+1$-dimensional wave equation with a positive, time-independent potential, the existence of exponentially growing modes is immediately ruled out and stability in this sense follows. With a little bit more of work, and based on the techniques described in Refs.~\cite{bKrW87} and ~\cite{mDiR08}, one can also prove that the solutions to the Regge-Wheeler and Zerilli equations belonging to sufficiently regular initial data on a spacelike slice remain uniformly bounded outside the black hole. For recent results on decay, see Refs.~\cite{mDiR08,mDiR09,rDwSaS11,rDwSaS12}.

The derivation of the Regge-Wheeler and Zerilli master equations was later clarified by the work of Moncrief~\cite{vM74a} who analyzed the gravitational perturbations from a Hamiltonian point of view and casted the Regge-Wheeler and Zerilli equations into gauge-invariant form. Moncrief also obtained master equations describing linear perturbations of spherically symmetric fluid stars~\cite{vM74b} and linear gravito-electromagnetic perturbations of a Reissner-Nordstr\"om black hole~\cite{vM74c,vM74d,vM75}, and showed that the latter is linearly stable in Einstein-Maxwell theory.\footnote{Interestingly, the magnetically charged Reissner-Nordstr\"om black hole is linearly {\em unstable} in Einstein-Yang-Mills-Higgs theory~\cite{kLvNeW92}.}

Later, Gerlach and Sengupta~\cite{uGuS79} provided a covariant description of the gauge-invariant perturbation approach, where instead of assuming a Schwarzschild background written in the usual Schwarzschild coordinates $(t,r)$ and foliated by the static $t=const$ slices, they only assumed the background to be spherically symmetric. Based on the natural $2+2$ form of the background induced by the metric two-spheres and the two-dimensional orbit manifold \hbox{$\tilde{M} = M\slash SO(3)$} orthogonal to them, and based on the decomposition into spherical tensor harmonics, Gerlach and Sengupta introduced a complete set of gauge-invariant tensor fields on the manifold $\tilde{M}$. This covariant approach has several advantages. Besides the possibility of describing perturbations of {\em dynamical} spacetimes, such as collapsing spherical stars, it also allows to describe the propagation of linearized gravitational waves on a Schwarzschild black hole in local coordinates which are {\em regular} at the event horizon, such as Kruskal- or ingoing Eddington-Finkelstein coordinates. Based on their approach, Gerlach and Sengupta obtained the covariant form of the Regge-Wheeler master equation, valid in any coordinate system compatible with the $2+2$ form of the background. The covariant form of the Zerilli master equation was derived in~\cite{oSmT01} (see also~\cite{oS00}) and a generalization including source terms with applications to the problem of calculating gravitational waves produced by the motion of a small star moving around a black hole was given in Ref.~\cite{kMeP05}. The covariant, gauge-invariant approach has also been applied to the derivation of master equations describing linear perturbations of black holes in general relativity coupled to a nonlinear electromagnetic theory, see Ref.~\cite{cMoS03} where sufficient conditions for the linear stability of such holes are also given. Further developments of the covariant gauge-invariant formalism include the coupling to a perfect fluid~\cite{cGjM00} and the generalization to second and higher-order perturbation theory~\cite{dBjMgM06,dBjMgM07,dBjMmT09,dBjMuSkK10}.

In a somewhat different development, linear perturbations of a Schwarzschild black hole were also analyzed in~\cite{jJ99} based on the $3+1$ formulation. The main advantage of this work is that, unlike previous approaches, the decomposition into spherical tensor harmonics is not needed, simplifying the derivation of the equations. Instead, two canonical pairs of gauge-invariant scalars are constructed, describing axial and polar perturbations in the "mono-dipole-free" sector (i.e. those fields corresponding to $\ell\geq 2$ in the decompositions into spherical harmonics). In the polar case, these gauge-invariant scalars involve \emph{quasilocal} operators which are local on $\tilde{M}$, but non-local on the two-spheres $S^2$. Based on these quantities, two scalar wave equations are derived which reduce to the covariant forms of the Regge-Wheeler and Zerilli master equations when decomposed into spherical harmonics. While the two scalar equations are presented in the covariant description, their derivation is based on the Schwarzschild coordinate patch.

Yet a different approach to gravitational perturbation theory which does not require the background to be spherically symmetric but assumes instead that it is {\em static} casts the perturbation equations into a wave equation for the linearized extrinsic curvature tensor~\cite{aAaAcL98,oBmHoS00,oSmHoB01,oS00}. This curvature-based approach has turned out to be useful for establishing the linear stability of certain Einstein-Yang-Mills black holes with a negative cosmological constant~\cite{oSeW01,eWoS02}.

In this work, we combine the covariant, gauge-invariant approach of~\cite{uGuS79} with the quasilocal method in~\cite{jJ99} and present a covariant, gauge-invariant perturbation formalism for an arbitrary spherically symmetric background without performing the expansion into spherical tensor harmonics. To this purpose, we  first review the relevant background equations describing the most general spherically symmetric spacetime in Sec.~\ref{Sec:Back}. Then, as a warmup, we derive in Sec.~\ref{Sec:Fields} master equations describing the propagation of scalar and electromagnetic test fields on such spacetimes. These master equations have the form of a wave equation on $\tilde{M}$ with an effective potential and act on an angular-dependent scalar field on $\tilde{M}$ which is gauge-invariant in the electromagnetic case. Next, in Sec.~\ref{Sec:Pert} we discuss linear metric perturbation and first review their behavior under infinitesimal coordinate transformations. Based on the quasilocal approach and a decomposition of tensor fields on the sphere in terms of scalars which is discussed in the appendix, we construct a set of gauge-invariant, angular-dependent tensor fields on $\tilde{M}$ which behave as scalars under rotations of the two-spheres. When performing an expansion into spherical harmonics, they reduce to the gauge-invariant tensors introduced by Gerlach and Sengupta. Then, we derive the expressions for the linearized Riemann curvature, Ricci and Einstein tensors, and obtain the perturbed Einstein equations. Next, in Sec.~\ref{Sec:Vac} we consider the vacuum case and derive the gauge-invariant master wave equations which reduce to the covariant Regge-Wheeler and Zerilli equations when an expansion in spherical harmonics is performed. We also show that the Regge-Wheeler equation can naturally be obtained in both the odd- \emph{and the even-parity} sectors and discuss how the metric perturbations can be reconstructed from the scalar potentials satisfying the master equations.

Next, in Sec.~\ref{Sec:Mat} we consider the coupling of the metric fields to arbitrary matter fields. In the odd-parity sector, we discuss the generalization of the Regge-Wheeler equation and state assumptions under which it yields a master equation. In the even-parity sector, we do not derive master equations for the metric fields; instead, we derive a constrained wave system for two gauge-invariant scalars from which the metric perturbations can be reconstructed under certain assumptions on the matter fields which should be reasonable. In the vacuum case, the two wave equations decouple from each other and are related to the radial parts of the Teukolsky equations~\cite{sT72} for the two Weyl scalars $\Psi_s$ with spin weights $s=-2$ and $s=2$, respectively. In fact, as we explain, our equations are equivalent to the ones obtained by Bardeen and Press~\cite{jBwP73}. Finally, in Sec.~\ref{Sec:Fluid}, we apply our formalism to the perturbations of self-gravitating spherical fluid configurations. Focussing on the fluid perturbations first, we decompose them into two parts, where the first part determines the perturbed vorticity and entropy and the second part describes a perturbed potential flow. The propagation of the first part is described by simple advection equations along the background flow. It decouples completely from the second part and the metric perturbations. Therefore, it can be solved separately. The second part couples to the metric perturbations, and together they obey and effective wave system on the orbit manifold $\tilde{M}$. This system should be useful for analytic and numerical investigations of the aforementioned problems. A summary and conclusions are given in Sec.~\ref{Sec:Conclusions}.

We use the signature convention $(-,+,+,+)$ for the metric and choose units for which $c=1$.

%%%%%%%%%%%%%%%%%%%%%%%%%%%%%%%%%%%%%%%%%%%
\section{Background equations}
\label{Sec:Back}
%%%%%%%%%%%%%%%%%%%%%%%%%%%%%%%%%%%%%%%%%%%

A spherically symmetric spacetime $(M,{\bf g})$ can be written as the product of a two-dimensional pseudo-Riemannian manifold $(\tilde{M},\tilde{{\bf g}})$ with the two-sphere $(S^2,\hat{{\bf g}})$,
\begin{equation}
M = \tilde{M} \times S^2,\qquad
{\bf g} = \tilde{g}_{ab} dx^a dx^b + r^2\hat{g}_{AB} dx^A dx^B,
\label{Eq:SphSymMetric}
\end{equation}
where $r$ is a strictly positive function on $\tilde{M}$, and $x^a,x^b,\ldots$ and $x^A,x^B,\ldots$ denote local coordinates on $\tilde{M}$ and $S^2$, respectively. The geometric interpretation of the function $r$ is the following: let $p\in \tilde{M}$ and consider the two-sphere $S_p := \{ p \} \times S^2\subset M$. Let $A(p) = |S_p|$ denote the area of $S_p$, computed from the induced metric on $S_p$. Therefore, $A(p) = 4\pi r(p)^2$ and the function $r$ is defined geometrically as
\begin{equation*}
r(p) = \sqrt{\frac{A(p)}{4\pi}}
\end{equation*}
and is called the areal radius. For the following, we assume $\tilde{M}$ to be oriented with volume form $\tilde{\varepsilon}_{ab} := \sqrt{|\tilde{g}|}\, \epsilon_{ab}$, where $|\tilde{g}| := |\det(\tilde{g}_{ab})|$ denotes the absolute value of the determinant of $\tilde{g}_{ab}$ and $\epsilon_{00} = \epsilon_{11} = 0$, $\epsilon_{01} = -\epsilon_{10} = 1$. We also introduce the differential $r_a := \tilde{\nabla}_a r$ and the covariant Hessian $r_{ab} := \tilde{\nabla}_a\tilde{\nabla}_b r$ of $r$ with respect to the covariant derivative $\tilde{\nabla}$ associated to the two-metric $\tilde{\bf {\bf g}}$.

The Christoffel symbols corresponding to the metric in Eq.~(\ref{Eq:SphSymMetric}) are
\begin{subequations}
\label{Eq:ChristoffelBackgr}
\begin{eqnarray}
&& \Gamma^d{}_{ab} = \tilde{\Gamma}^d{}_{ab}\, ,\qquad
   \Gamma^d{}_{aB} = 0, \qquad
   \Gamma^d{}_{AB} = -r r^d \hat{g}_{AB}\, ,\\
&& \Gamma^D{}_{ab} = 0,\qquad
   \Gamma^D{}_{aB} = \frac{r_a}{r}\,\delta^D{}_B\, , \qquad
   \Gamma^D{}_{AB} = \hat{\Gamma}^D{}_{AB}\, ,
\end{eqnarray}
\end{subequations}
where here and in the following, quantities with a tilde and a hat refer to the manifolds $(\tilde{M},\tilde{\bf g})$ and $(S,\hat{\bf g})$, respectively. From this, one finds the following expressions for the curvature tensor,
\begin{subequations}
\label{Eq:CurvatureBackgr}
\begin{eqnarray}
R_{abcd} &=& \tilde{R}_{abcd} 
 = -\tilde{k}\tilde{\varepsilon}_{ab}\tilde{\varepsilon}_{cd}
 = \tilde{k}(\tilde{g}_{ac}\tilde{g}_{bd} - \tilde{g}_{ad}\tilde{g}_{bc}),\\
R_{aBcd} &=& 0,\\
R_{aBcD} &=& -r r_{ac} \hat{g}_{BD},\\
R_{abCD} &=& 0,\\
R_{aBCD} &=& 0,\\
R_{ABCD} &=& r^2(1 - N)(\hat{g}_{AC}\hat{g}_{DB} - \hat{g}_{AD}\hat{g}_{CB}),
\end{eqnarray}
\end{subequations}
where $\tilde{k}$ denotes the Gauss curvature of $(\tilde{M},\tilde{{\bf g}})$ and where $N := \tilde{{\bf g}}(dr,dr) = r^a r_a$. Contracting, one obtains the Ricci tensor,
\begin{subequations}
\begin{eqnarray}
R_{ab} &=& \tilde{k}\tilde{g}_{ab} - 2\frac{r_{ab}}{r},\\
R_{aB} &=& 0,\\
R_{AB} &=& \left( 1 - N - r\tilde{\Delta} r \right)\hat{g}_{AB},
\end{eqnarray}
\label{Eq:RicciBackgr}
\end{subequations}
where $\tilde{\Delta} r := \tilde{g}^{ab} r_{ab} = \tilde{\nabla}^a\tilde{\nabla}_a r$. From this, one finally obtains the Einstein tensor,
\begin{subequations}
\label{Eq:EinsteinBackgr}
\begin{eqnarray}
G_{ab} &=& -\frac{2}{r}(r_{ab})^{tf} 
 - \frac{1}{r^2}\tilde{g}_{ab}\left( 1 - N - r\tilde{\Delta} r \right),\\
G_{aB} &=& 0,\\
G_{AB} &=& \left( r\tilde{\Delta} r - \tilde{k} r^2 \right)\hat{g}_{AB},
\end{eqnarray}
\end{subequations}
where $(r_{ab})^{tf} := r_{ab} - \frac{1}{2}\tilde{g}_{ab}\tilde{g}^{cd} r_{cd}$ denotes the trace-free part of $r_{ab}$. This particular structure of the Einstein tensor implies that the stress-energy tensor ${\bf T}$ must satisfy the conditions $T_{aB} = 0$ and $T_{AB}$ proportional to $\hat{g}_{AB}$. Summarizing, Einstein's field equations in spherical symmetry consist of
\begin{subequations}
\label{Eq:EinsteinSph}
\begin{eqnarray}
&& -\frac{2}{r}(r_{ab})^{tf} = \kappa T_{ab}^{tf},
\label{Eq:EinsteinSph1}\\
&& -\frac{2}{r^2} \left(1 - N - r\tilde{\Delta} r \right) = \kappa \tilde{g}^{ab} T_{ab},
\label{Eq:EinsteinSph2}\\
&& 2\left( r\tilde{\Delta} r - \tilde{k} r^2 \right) = \kappa\hat{g}^{AB} T_{AB},
\label{Eq:EinsteinSph3}
\end{eqnarray}
\end{subequations}
where $\kappa := 8\pi G_N$ with Newton's constant $G_N$.

The Bianchi identities imply that
\begin{equation*}
0 = \nabla_\mu G^\mu{}_b 
  = \frac{1}{r^2}\tilde{\nabla}_a( r^2 G^a{}_b ) - \frac{r_b}{r^3}\hat{g}^{AB} G_{AB} .
\end{equation*}
If the stress-energy tensor is divergence-free, the same equation holds for ${\bf T}$, and hence,
\begin{equation*}
\frac{r_b}{r}\hat{g}^{AB}( G_{AB} - \kappa T_{AB} ) 
 =  \tilde{\nabla}_a\left[ r^2 (G^a{}_b - \kappa T^a{}_b) \right]. 
\end{equation*}
This shows that Eq.~(\ref{Eq:EinsteinSph3}) follows from Eqs.~(\ref{Eq:EinsteinSph1},\ref{Eq:EinsteinSph2}) provided $r_b\neq 0$ and $\nabla_\mu T^{\mu\nu} = 0$ hold, and in this case it is sufficient to solve Eqs.~(\ref{Eq:EinsteinSph1},\ref{Eq:EinsteinSph2}).

%%%%%%%%%%%%%%%%%%%%%%%%%%%%%%%%%%%%%%%%%%%%%
\section{Scalar and electromagnetic fields propagating on a spherically symmetric background}
\label{Sec:Fields}
%%%%%%%%%%%%%%%%%%%%%%%%%%%%%%%%%%%%%%%%%%%%%

In this section we derive the master equations describing scalar and electromagnetic fields propagating on an arbitrary spherically symmetric spacetime.

\subsection{Scalar field propagation}
\label{Sec:FieldsScalar}

The propagation of a scalar field on a fixed spacetime background $(M,{\bf g})$ is described by Klein-Gordon equation
\begin{equation}
(\Box_{\bf g} + \mu^2)\phi = 0,
\label{Eq:KG}
\end{equation}
where $\phi$ is a real scalar field on $M$ with mass $\mu$ (in units for which $\hbar=c=1)$, and $\Box_{\bf g} := -\nabla^\alpha\nabla_\alpha$ is the covariant d'Alembertian on $(M,{\bf g})$. Assuming that $(M,{\bf g})$ is spherically symmetric and applying the general formula 
\begin{equation*}
\Box_{\bf g}\phi = -\frac{1}{\sqrt{|g|}}\frac{\partial}{\partial x^\alpha}\left(
 \sqrt{|g|} g^{\alpha\beta}\frac{\partial\phi}{\partial x^\beta} \right),\qquad
 |g| := |\det(g_{\alpha\beta})|,
\end{equation*}
to a local coordinate patch, we can rewrite Eq.~(\ref{Eq:KG}) in its $2+2$ form,
\begin{equation*}
-\frac{1}{r^2}\tilde{\nabla}^a( r^2\tilde{\nabla}_a\phi ) 
+ \left( -\frac{\hat{\Delta}}{r^2} + \mu^2 \right)\phi = 0,
\end{equation*}
where $\hat{\Delta} = \hat{g}^{AB}\hat{\nabla}_A\hat{\nabla}_B$ is the Laplacian on the round sphere $(S^2,\hat{\bf g})$. Finally, we introduce the rescaled scalar field $\psi := r\phi$ and use the identity
\begin{equation*}
-\tilde{\nabla}^a(r^2\tilde{\nabla}_a\phi) = r\tilde{\Box}\psi + (\tilde{\Delta} r)\psi,
\end{equation*}
with $\tilde{\Box} := -\tilde{\Delta} = -\tilde{\nabla}^a\tilde{\nabla}_a$ the covariant d'Alembertian on $(\tilde{M},\tilde{\bf g})$. This yields
\begin{equation}
\tilde{\Box}\psi + \left( -\frac{\hat{\Delta}}{r^2}
 + \frac{\tilde{\Delta} r}{r} + \mu^2 \right)\psi = 0.
\label{Eq:EffectiveKG}
\end{equation}
This has the form of a wave equation on $(\tilde{M},\tilde{{\bf g}})$ with effective potential
\begin{equation*}
V := -\frac{\hat{\Delta}}{r^2} + \frac{\tilde{\Delta} r}{r} + \mu^2.
\end{equation*}
Here, the first term is an operator on the sphere $S^2$. If $\psi$ is decomposed into spherical harmonics, it becomes the multiplicative operator $\ell(\ell+1)/r^2$ with $\ell$ the angular momentum number, which represents the usual centrifugal term. The second term is a curvature correction term which can be further simplified using the Einstein equation~(\ref{Eq:EinsteinSph2}). For a Schwarzschild spacetime of mass $m$, for instance, for which $N = 1 - 2m/r$, we have $r^{-1}\tilde{\Delta} r = 2m/r^3$. The third term in the potential, $\mu^2$, is just inherited from the corresponding term of the original Klein-Gordon equation~(\ref{Eq:KG}).

As we will see below, the effective equations describing electromagnetic and linearized fluctuations on spherically symmetric spacetimes have a form very similar to Eq.~(\ref{Eq:EffectiveKG}).

\subsection{Electromagnetic propagation}

Next, we consider Maxwell's equations
\begin{equation}
-\nabla^\mu F_{\mu\nu} = J_\nu,\qquad
F_{\mu\nu} = \nabla_\mu A_\nu - \nabla_\nu A_\mu
\label{Eq:Maxwell}
\end{equation}
on a spherically symmetric background $(M,{\bf g})$. Here, ${\bf A} = A_\mu dx^\mu$ is the electromagnetic potential one-form, ${\bf F} := d{\bf A} = 2^{-1} F_{\mu\nu} dx^\mu\wedge dx^\nu$ the corresponding Faraday tensor, and ${\bf J} = J_\mu dx^\mu$ is the electric four-current density.

According to the $2+2$ form of the background, we may decompose the electromagnetic potential as
\begin{equation*}
{\bf A} = \alpha_a dx^a + \beta_B dx^B,
\end{equation*}
where the quantities $\alpha_a$ and $\beta_B$ both depend on the coordinates $(x^a, x^B)$. Notice that with respect to a rotation on $S^2$, $\beta_A$ transforms like the components of a one-form while $\alpha_a$ transforms as a scalar. With respect to a gauge-transformation ${\bf A} \mapsto {\bf A} + d\xi$ parametrized by a function $\xi$ on $M$, we have
\begin{equation}
\alpha_a \mapsto \alpha_a + \tilde{\nabla}_a\xi,\qquad
\beta_B \mapsto \beta_B + \hat{\nabla}_B\xi.
\label{Eq:MaxwellGT}
\end{equation}
Quantities which are invariant with respect to these transformations can be constructed based on the following decomposition for a one-form ${\bm\omega} =\omega_A dx^A$ in terms of two scalar fields $f$ and $g$ on the two-sphere $S^2$:
\begin{equation}
\omega_A = \hat{\nabla}_A f + \hat{\varepsilon}_A{}^B\hat{\nabla}_B g,
\label{Eq:1FormDecomp}
\end{equation}
where $\hat{\varepsilon}_{AB}$ denotes the volume form on $S^2$. See the appendix for a proof and further discussion on this decomposition. Here, we note that the two terms on the right-hand side of Eq.~(\ref{Eq:1FormDecomp}) are mutually orthogonal with respect to the natural $L^2$ scalar product,
\begin{equation*}
\langle {\bm\omega},{\bm\eta} \rangle 
 := \int\limits_{S^2} \hat{g}^{AB}\omega_A\eta_B \sqrt{|\hat{g}|} d^2 x
\end{equation*}
for two one-forms ${\bm\omega}$ and ${\bm\eta}$ on $S^2$. Therefore, the decomposition~(\ref{Eq:1FormDecomp}) is unique. However, the functions $f$ and $g$ themselves are only unique up to an additive constant. In the following, we fix this constant by requiring $f$ and $g$ to have zero mean values over $(S^2,\hat{\bf g})$.

The decomposition~(\ref{Eq:1FormDecomp}) allows us to represent
\begin{equation*}
\beta_B = \hat{\nabla}_B \mu + \hat{\varepsilon}_B{}^C\hat{\nabla}_C \nu
\end{equation*}
in terms of two scalar fields $\mu$ and $\nu$. Since $\beta_B$ depends not only on the angular variables, but also on the radial ones, the fields $\mu$ and $\nu$ depend on both $x^a$ and $x^B$. However, unlike $\beta_B$, they transform like scalar fields under rotations of the two-spheres. Similarly, the one-form $\alpha_a dx^a$ depends on the angular coordinates $x^B$, but it transforms as a scalar field under rotations. As mentioned above, from now on we assume that $\mu$ and $\nu$ have zero mean values over $(S^2,\hat{\bf g})$. This implies that their decompositions into spherical harmonics have a vanishing monopole term.

Working in the monopole-free space, the gauge transformations~(\ref{Eq:MaxwellGT}) imply the following transformations for $\alpha_a$, $\mu$ and $\nu$:
\begin{equation}
\alpha_a \mapsto \alpha_a + \tilde{\nabla}_a\xi,\qquad
\mu\mapsto\mu + \xi,\qquad
\nu\mapsto\nu.
\end{equation}
Therefore, the following quantities are {\em gauge-invariant}, i.e. invariant with respect to the transformations~(\ref{Eq:MaxwellGT}):
\begin{equation}
\alpha_a^{(inv)} := \alpha_a - \tilde{\nabla}_a\mu,\qquad
\nu^{(inv)} := \nu.
\end{equation}

The Faraday tensor ${\bf F}$ is gauge-invariant, and hence it must be possible to express it solely in terms of $\alpha_a^{(inv)}$ and $\nu^{(inv)}$. A short calculation reveals that
\begin{equation*}
F_{ab} = \tilde{\nabla}_a\alpha_b^{(inv)} - \tilde{\nabla}_b\alpha_a^{(inv)},\qquad
F_{aB} = -F_{Ba} = -\hat{\nabla}_B\alpha_a^{(inv)}
 + \hat{\varepsilon}_B{}^C\hat{\nabla}_C\tilde{\nabla}_a\nu^{(inv)},\qquad
F_{AB} = -\hat{\varepsilon}_{AB}\hat{\Delta} \nu^{(inv)}.
\end{equation*}
Applying the general formula ($F^{\mu\nu} = -F^{\nu\mu}$)
\begin{equation*}
\nabla_\mu F^{\mu\nu} = \frac{1}{\sqrt{|g|}}\frac{\partial}{\partial x^\mu}
\left( \sqrt{|g|} F^{\mu\nu} \right)
\end{equation*}
to a coordinate patch, we obtain the following Maxwell equations in their $2+2$ form,
\begin{subequations}
\label{Eq:MaxwellEvenOdd}
\begin{eqnarray}
-\tilde{\nabla}^a\left[ r^2\left( 
 \tilde{\nabla}_a\alpha_b^{(inv)} - \tilde{\nabla}_b\alpha_a^{(inv)} \right) \right]
 - \hat{\Delta}\alpha_b^{(inv)} &=& r^2 J_b,
\label{Eq:MaxwellEven1}\\
\tilde{\nabla}^a\alpha_a^{(inv)} &=& j,
\label{Eq:MaxwellEven2}\\
\tilde{\Box}\nu^{(inv)} - \frac{1}{r^2}\hat{\Delta}\nu^{(inv)} &=& k,
\label{Eq:MaxwellOdd}
\end{eqnarray}
\end{subequations}
where we have decomposed the angular components of the four-current according to
$J_B = \hat{\nabla}_B j + \hat{\varepsilon}_B{}^C\hat{\nabla}_C k$. Eq.~(\ref{Eq:MaxwellOdd}) decouples from the remaining two equations, and has the form of an inhomogeneous wave equation on $(\tilde{M},\tilde{\bf g})$ with effective potential $V = -r^{-2}\hat{\Delta}$. This master equation describes the propagation of the odd-parity (or axial) part of the electromagnetic field.

In order to derive a master equation for the even-parity (or polar) part of the field, we first introduce the scalar
\begin{equation*}
\Phi := r^2\tilde{\varepsilon}^{ab}\tilde{\nabla}_a\alpha_b^{(inv)},
\end{equation*}
in terms of which $F_{ab} = -r^{-2}\tilde{\varepsilon}_{ab}\Phi$ and Eq.~(\ref{Eq:MaxwellEven1}) can be rewritten as
\begin{equation}
\tilde{\varepsilon}_{ab}\tilde{\nabla}^a\Phi - \hat{\Delta}\alpha_b^{(inv)} = r^2 J_b.
\label{Eq:MaxwellEvenInt}
\end{equation}
Applying the two-dimensional curl operator $\tilde{\varepsilon}^{ab}\tilde{\nabla}_a$ on both sides of this equation we obtain
\begin{equation}
\tilde{\Box}\Phi - \frac{1}{r^2}\hat{\Delta}\Phi 
= \tilde{\varepsilon}^{ab}\tilde{\nabla}_a( r^2 J_b ),
\label{Eq:MaxwellMasterEq}
\end{equation}
which is an inhomogeneous wave equation on $(\tilde{M},\tilde{{\bf g}})$ of exactly the same form as the odd-parity equation~(\ref{Eq:MaxwellOdd}). It is a master equation describing the propagation of the even-parity part of the field: solving Eq.~(\ref{Eq:MaxwellMasterEq}) gives $\Phi$, from which the gauge-invariant quantity $\alpha_a^{(inv)}$ can be reconstructed from Eq.~(\ref{Eq:MaxwellEvenInt}) since the Laplacian $\hat{\Delta}$ is invertible on the monopole-free space. This quantity automatically satisfies Eq.~(\ref{Eq:MaxwellEven2}) since by Eq.~(\ref{Eq:MaxwellEvenInt}) and the continuity equation $\nabla_\mu J^\mu = 0$ we have $-\hat{\Delta}(\tilde{\nabla}^b\alpha_b^{(inv)}) = \tilde{\nabla}^b(r^2 J_b) = -\hat{\Delta} j$. Therefore, we can reconstruct the even-parity part of the Faraday tensor, and all the Maxwell equations are satisfied.

%%%%%%%%%%%%%%%%%%%%%%%%%%%%%%%%%%%%%%%%%%%%%
\section{Gravitational perturbation equations}
\label{Sec:Pert}
%%%%%%%%%%%%%%%%%%%%%%%%%%%%%%%%%%%%%%%%%%%%%

In this section we describe linearized metric perturbations on a spherically symmetric background. We start by analyzing their behavior under an infinitesimal coordinate change, and based on ideas introduced in the previous section, construct a full set of gauge-invariant, angular-dependent tensor fields on $\tilde{M}$. Similar invariants are also constructed from the perturbed stress-energy tensor. Then, we derive the linearized Einstein equations and write them in terms of these gauge-invariant quantities.

We consider a smooth perturbation of the spherically symmetric metric ${\bf g}$, that is, a smooth, one-parameter family\footnote{It is also possible to perturb $M$, considering a one-parameter family of spacetimes $(M(\lambda),{\bf g}(\lambda))$. In this case, the manifolds $M(\lambda)$ need to be identified with each other by an appropriate diffeomorphism, which leads to an alternative, but equivalent point of view. For a discussion which is based on this approach and analyzes the behavior of linear and nonlinear perturbations under diffeomorphisms, see Ref.~\cite{mBsMsMsS97} for instance.}  of metrics, ${\bf g}(\lambda)$, on $M$ such that ${\bf g}(0) = {\bf g}$. To first order, the deviation from the background metric ${\bf g}$ is described by the variation
\begin{equation*}
\delta {\bf g} := \left. \frac{d}{d\lambda} {\bf g}(\lambda) \right|_{\lambda = 0},
\end{equation*}
whose geometric interpretation is the tangent vector to the curve ${\bf g}(\lambda)$ at the point ${\bf g} = {\bf g}(0)$ in configuration space. With respect to a one-parameter family of diffeomorphisms, $\phi(\lambda) : M \to M$, the metrics ${\bf g}(\lambda)$ transform according to
\begin{equation}
{\bf g}(\lambda) \mapsto \phi(\lambda)^* {\bf g}(\lambda),
\label{Eq:Diffeo}
\end{equation}
the star denoting pull-back. We may split $\phi(\lambda) = \psi(\lambda)\varphi$, where $\varphi:=\phi(0): M \to M$ is the diffeomorphism to zeroth order in $\lambda$ and where $\psi(\lambda) : M \to M$ satisfies $\psi(0) = \mbox{id}$. Differentiating Eq.~(\ref{Eq:Diffeo}) with respect to $\lambda$, we then obtain
\begin{equation}
\delta {\bf g} \mapsto \varphi^*( \delta {\bf g} + \pounds_{\bf X} {\bf g} ),
\label{Eq:CoordTrans}
\end{equation}
where the vector field ${\bf X}$ is defined as the variation of $\psi$,
\begin{equation*}
{\bf X}_p = \left. \frac{d}{d\lambda} \psi(\lambda)(p) \right|_{\lambda = 0}.
\end{equation*}
For the following, we will restrict ourselves to "background" diffeomorpisms $\varphi$ which leave the structure $M = \tilde{M} \times S^2$ of the manifold invariant. The next step in the discussion of our perturbation formalism is to construct quantities from $\delta {\bf g}$ which are invariant with respect to the infinitesimal coordinate transformations $\delta {\bf g} \mapsto \delta {\bf g} + \pounds_{\bf X} {\bf g}$.

%%%%%%%%%%%%%%%%%%%%%%%%%%%%%%%%%%%%%%%%%%%%%
\subsection{$2+2$ split and gauge invariance}

For the following, it is convenient to split the metric perturbations $\delta {\bf g}$ in accordance with the $2+2$ form of the background metric:
\begin{equation*}
H_{ab} := \delta g_{ab}\; ,\qquad
Q_{aB} := \delta g_{aB}\; ,\qquad
K_{AB} := r^{-2}\delta g_{AB}\; ,
\end{equation*}
where the quantities $H_{ab}$, $Q_{aB}$ and $K_{AB}$ depend on the coordinates $(x^a,x^B)$. Similarly, we may split the vector field ${\bf X}$ in Eq.~(\ref{Eq:CoordTrans}) according to
\begin{equation*}
\xi^a := X^a,\qquad
\eta^B := X^B.
\end{equation*}
With respect to the infinitesimal coordinate transformation generated by ${\bf X}$ we then have
\begin{subequations}
\label{Eq:CoordTransf}
\begin{eqnarray}
H_{ab} &\mapsto& H_{ab} + 2\tilde{\nabla}_{(a} \xi_{b)}\; ,
\label{Eq:CoordTransf1}\\
Q_{aB} &\mapsto& Q_{aB} + \hat{\nabla}_B\xi_a 
 + r^2\tilde{\nabla}_a \hat{\eta}_B\, ,
\label{Eq:CoordTransf2}\\
K_{AB} &\mapsto& K_{AB} + \frac{2}{r}\,\hat{g}_{AB}\, r^a \xi_a
 + 2\hat{\nabla}_{(A} \hat{\eta}_{B)}\, ,
\label{Eq:CoordTransf3}
\end{eqnarray}
\end{subequations}
with $\hat{\eta}_A := \hat{g}_{AB}\eta^B$. 

In order to construct quantities which are invariant with respect to these transformations, we generalize the ideas presented in the previous section and decompose any tensor field on $S^2$ in terms of scalars. For one-forms ${\bm\omega}$ on $S^2$ we recall the decomposition in Eq.~(\ref{Eq:1FormDecomp}) above. For a symmetric, trace-free tensor field ${\bm\tau} = \tau_{AB} dx^A dx^B$ on $S^2$ we use the following decomposition proved in the appendix,
\begin{equation}
\tau_{AB} = (\hat{\nabla}_A\hat{\nabla}_B F)^{TF} 
 + \hat{\varepsilon}_{(A}{}^C\hat{\nabla}_{B)}\hat{\nabla}_C G,
\label{Eq:SymTensorDecomp}
\end{equation}
where $F$ and $G$ are scalar fields on $S^2$ and the super-index ${}^{TF}$ denotes  the trace-free part with respect to $\hat{\bf g}$. This decomposition is also orthogonal with respect to the natural $L^2$ scalar product on $S^2$ and therefore, it is unique. However, the functions $F$ and $G$ are only unique up to the addition of a monopole or dipole term (that is, a function $L$ on $S^2$ satisfying $(\hat{\nabla}_A\hat{\nabla}_B L)^{TF} = 0$). We fix this freedom by working on the mono-dipole-free space defined as the orthogonal complement of the monopole and dipole terms. 

Therefore, we may decompose the perturbations $Q_{aB}$ and $K_{AB}$ according to
\begin{subequations}
\begin{eqnarray}
Q_{aB} &=& \hat{\nabla}_B q_a + \hat{\varepsilon}_B{}^C\hat{\nabla}_C h_a,\\
K_{AB} &=& 2(\hat{\nabla}_A\hat{\nabla}_B G)^{TF} + \frac{1}{2}\hat{g}_{AB} J
 + 2\hat{\varepsilon}_{(A}{}^C\hat{\nabla}_{B)}\hat{\nabla}_C k,
\end{eqnarray}
\end{subequations}
where ${\bf q} = q_a dx^a$ and ${\bf h} = h_a dx^a$ are angular-dependent one-forms on $\tilde{M}$, and $G$, $J$ and $k$ are angular-dependent functions on $\tilde{M}$. The advantage of this decomposition relies in the fact that it is covariant. In particular, the perturbations are fully determined by quantities which transform like scalar fields with respect to diffeomorphisms on the sphere (i.e. a diffeomorphism $\varphi: M \to M$ which leaves each point of $\tilde{M}$ invariant). Here, $J = \hat{g}^{AB} K_{AB}$ represents the trace of $K_{AB}$. Similarly, we decompose $\hat{\eta}_B = \hat{\nabla}_B f + \hat{\varepsilon}_B{}^C\hat{\nabla}_C g$. In order to define ${\bf q}$, ${\bf h}$, $G$, $J$, $k$, $f$ and $g$ uniquely, we suppose that these quantities lie in the mono-dipole-free space in what follows.\footnote{For a detailed discussion on monopole and dipole perturbations in the vacuum case, see Refs.~\cite{jJ99,oSmT01}, where it is shown that they correspond to stationary modes, i.e. small changes in the mass or small rotations. For the Einstein-Euler case, see Ref.~\cite{cGjM00}.} As a consequence of our assumptions, the transformations~(\ref{Eq:CoordTransf}) induce the following transformations for the perturbation amplitudes ${\bf H}:=H_{ab} dx^a dx^b$, ${\bf q}$, ${\bf h}$, $G$, $J$, $k$:
\begin{subequations}
\begin{eqnarray}
H_{ab} &\mapsto& H_{ab} + \tilde{\nabla}_a\xi_b + \tilde{\nabla}_b\xi_a,\\
q_a &\mapsto& q_a + \xi_a + r^2\tilde{\nabla}_a f,\\
h_a &\mapsto& h_a + r^2\tilde{\nabla}_a g,\\
G &\mapsto& G + f,\\
J &\mapsto& J + \frac{4}{r}\, r^a\xi_a + 2\hat{\Delta} f,\\
k &\mapsto& k + g.
\end{eqnarray}
\end{subequations}
There are two commonly used methods for dealing with these transformations. 
The first, which is known as {\em gauge-fixing}, imposes conditions on the perturbations amplitudes which fix the gauge functions $\xi_a$, $f$ and $g$. A simple way of achieving this is to demand $k=0$, $G=0$, and $q_a=0$, which simplifies the perturbations considerably. This gauge is called the {\em Regge-Wheeler gauge}. The second method, called the {\em gauge-invariant approach}, does not impose any conditions on the perturbation amplitudes. Instead, one constructs linear combinations of the perturbation amplitudes which are invariant with respect to the transformations above. For example, we may define the gauge-invariant one-form
\begin{equation}
h_a^{(inv)} := h_a - r^2\tilde{\nabla}_a k,
\end{equation}
which transforms trivially, $h_a^{(inv)}\mapsto h_a^{(inv)}$. The remaining gauge-invariants are obtained by first noticing that $p_a := q_a - r^2\tilde{\nabla}_a G$ transforms like $p_a \mapsto p_a + \xi_a$, and then setting
\begin{subequations}
\begin{eqnarray}
H^{(inv)}_{ab} &:=& H_{ab} -  \tilde{\nabla}_a p_b - \tilde{\nabla}_b p_a,\\
J^{(inv)} &:=& J - \frac{4}{r}\, r^a p_a - 2\hat{\Delta} G.
\end{eqnarray}
\end{subequations}
The advantage of the gauge-invariant approach is that no gauge conditions need to be imposed. Therefore, one does not need to worry about the physical results obtained depending on a specific coordinate choice. Since the field equations are gauge-invariant, it is clear that the linearized field equations can be expressed in terms of such gauge-invariant quantities only.

The variation of the stress-energy tensor, $\delta {\bf T}$, may be decomposed similarly to the variation of the metric:
\begin{subequations}
\label{Eq:LinTmunu}
\begin{eqnarray}
\delta T_{ab} &=& \tau_{ab},
\label{Eq:LinTab}\\
\delta T_{aB} &=& \hat{\nabla}_B\mu_a + \hat{\varepsilon}_B{}^C\hat{\nabla}_C\nu_a,
\label{Eq:LinTaB}\\
\delta T_{AB} &=& r^2\left[ 2(\hat{\nabla}_A\hat{\nabla}_B \alpha)^{TF} 
 + \frac{1}{2}\hat{g}_{AB} \lambda
 + 2\hat{\varepsilon}_{(A}{}^C\hat{\nabla}_{B)}\hat{\nabla}_C\beta \right].
\label{Eq:LinTAB}
\end{eqnarray}
\end{subequations}
With respect to an infinitesimal coordinate transformation generated by the vector field ${\bf X}$ we have $\delta {\bf T} \mapsto \delta {\bf T} + \pounds_{\bf X} {\bf T}$. Explicitly, this gives
\begin{subequations}
\begin{eqnarray}
\tau_{ab} &\mapsto& \tau _{ab} + \tilde{\pounds}_\xi T_{ab},\\
\mu_a &\mapsto& \mu_a + T_{ab}\xi^b + r^2 P\tilde{\nabla}_a f,\\
\nu_a &\mapsto& \nu_a + r^2 P\tilde{\nabla}_a g,\\
\alpha &\mapsto& \alpha + P f,\\
\lambda &\mapsto& \lambda + \frac{2}{r^2}\tilde{\nabla}^a(r^2 P)\xi_a 
 + 2P \hat{\Delta} f,\\
\beta &\mapsto& \beta + P g,
\end{eqnarray}
\end{subequations}
where we have used $T_{AB} = r^2 P\hat{g}_{AB}$ with $P:=\hat{g}^{AB} T_{AB}/(2r^2)$. From this, we can construct the following gauge invariants:
\begin{subequations}
\begin{eqnarray}
\tau_{ab}^{(inv)}  &:=& \tau _{ab} 
 - p^c\tilde{\nabla}_c T_{ab} - 2 T_{c(a}\tilde{\nabla}_{b)} p^c,\\
\mu_a^{(inv)}  &:=& \mu_a - T_{ab} p^b - r^2 P\tilde{\nabla}_a G,\\
\nu_a^{(inv)}  &:=& \nu_a - r^2 P\tilde{\nabla}_a k,\\
\alpha^{(inv)}  &:=& \alpha - P G,\\
\lambda^{(inv)}  &:=& \lambda - \frac{2}{r^2}\tilde{\nabla}^a(r^2 P) p_a - 2P\hat{\Delta} G,\\
\beta^{(inv)}  &:=& \beta - P k.
\end{eqnarray}
\end{subequations}
For the explicit calculations below, the following trick will be used: as explained above, in the Regge-Wheeler gauge, the perturbations simplify considerably since $q_a=0$ and $G=k=0$. Furthermore, in this particular gauge, we have $h^{(inv)}_a = h_a$, $H^{(inv)}_{ab} = H_{ab}$ and $J^{(inv)} = J$, and $\tau_{ab}^{(inv)} = \tau_{ab}$, $\mu_a^{(inv)} = \mu_a$ etc. Since the field equations are gauge-invariant, it is sufficient to perform the calculations in this special gauge; the results in an arbitrary gauge can be obtained by simply replacing $h_a$ by $h_a^{(inv)}$, $H_{ab}$ by $H_{ab}^{(inv)}$, $J$ by $J^{(inv)}$, $\tau_{ab}$ by $\tau_{ab}^{(inv)}$ etc. in the final equations.

%%%%%%%%%%%%%%%%%%%%%%%%%%%%%%%%%%%%%%%%%%%%%
\subsection{Computation of the linearized Einstein equations}

As we have just discussed, it is sufficient to consider the following form for the variation of the metric,
\begin{eqnarray*}
\delta g_{ab} &=& H_{ab},\\
\delta g_{aB} &=& \hat{\varepsilon}_B{}^C\hat{\nabla}_C h_a,\\
\delta g_{AB} &=& \frac{1}{2} r^2\hat{g}_{AB} J.
\end{eqnarray*}
With respect to this $2+2$ decomposition the linearized Christoffel symbols,
\begin{eqnarray*}
\delta\Gamma^\mu{}_{\alpha\beta} &=& \frac{1}{2}\, g^{\mu\nu}
\left( \partial_\alpha \delta g_{\beta\nu} + \partial_\beta \delta g_{\alpha\nu} 
 - \partial_\nu \delta g_{\alpha\beta} 
 - 2\Gamma^\sigma{}_{\alpha\beta}\delta g_{\sigma\nu} \right) \\
 &=& \frac{1}{2}\, g^{\mu\nu}
\left( \nabla_\alpha \delta g_{\beta\nu} + \nabla_\beta \delta g_{\alpha\nu} 
 - \nabla_\nu \delta g_{\alpha\beta} \right),
\end{eqnarray*}
are
\begin{subequations}
\label{Eq:LinChristoffel}
\begin{eqnarray}
\delta\Gamma^c{}_{ab} &=& \frac{1}{2}\, \tilde{g}^{cd}
\left( \tilde{\nabla}_a H_{bd} + \tilde{\nabla}_b H_{ad} 
 - \tilde{\nabla}_d H_{ab} \right),
\label{Eq:LinChristoffel1}\\
\delta\Gamma^c{}_{aB} &=& \frac{1}{2}\, \tilde{g}^{cd}
\left[ \hat{\nabla}_B H_{ad} + \hat{\varepsilon}_B{}^C\hat{\nabla}_C\left(
\tilde{\nabla}_a h_d - \tilde{\nabla}_d h_a - 2\frac{r_a}{r}\, h_d \right) \right],
\\
\delta\Gamma^c{}_{AB} &=& \tilde{g}^{cd} 
\left[ \hat{g}_{AB} r r^a H_{ad} - \frac{1}{4}\hat{g}_{AB}\tilde{\nabla}_d(r^2 J)
 + \hat{\varepsilon}_{(A}{}^C\hat{\nabla}_{B)}\hat{\nabla}_C h_d \right],
\\
\delta\Gamma^C{}_{ab} &=& -\frac{1}{2r^2}\, \hat{g}^{CD}\hat{\nabla}_D H_{ab}
 + \hat{\varepsilon}^{CD}\hat{\nabla}_D \left( \frac{1}{r^2} \tilde{\nabla}_{(a} h_{b)} \right),
\\
\delta\Gamma^C{}_{aB} &=& \frac{1}{4}\delta^C{}_B\tilde{\nabla}_a J
 + \frac{1}{2r^2}\hat{\varepsilon}^C{}_B\hat{\Delta} h_a,
\\
\delta\Gamma^C{}_{AB} &=& \frac{1}{4}\left(
\delta^C{}_B\hat{\nabla}_A J + \delta^C{}_A\hat{\nabla}_B J 
 - \hat{g}_{AB}\hat{g}^{CD}\hat{\nabla}_D J \right) 
 + \hat{g}_{AB}\hat{\varepsilon}^{CD}\hat{\nabla}_D\left( \frac{r^a}{r} h_a \right).
\label{Eq:LinChristoffel6}
\end{eqnarray}
\end{subequations}

Next, we compute the linearized Riemann curvature tensor,
\begin{eqnarray}
\delta R^\mu{}_{\nu\alpha\beta} &=& \partial_\alpha \delta\Gamma^\mu{}_{\beta\nu}
 + \Gamma^\mu{}_{\alpha\sigma}\delta\Gamma^\sigma{}_{\beta\nu}
 + \Gamma^\sigma{}_{\beta\nu}\delta\Gamma^\mu{}_{\alpha\sigma}
 - (\alpha\leftrightarrow\beta)
\nonumber\\ 
&=& \nabla_\alpha \delta\Gamma^\mu{}_{\beta\nu} 
  - \nabla_\beta \delta\Gamma^\mu{}_{\alpha\nu}.
\nonumber
\end{eqnarray}
Using the expressions~(\ref{Eq:LinChristoffel}), the fact that $\delta R_{\mu\nu\alpha\beta} = \delta( g_{\mu\sigma} R^\sigma{}_{\nu\alpha\beta} ) = g_{\mu\sigma}\delta R^\sigma{}_{\nu\alpha\beta} + R^\sigma{}_{\nu\alpha\beta}\delta g_{\mu\sigma}$, and the symmetries of the curvature tensor, implying
\begin{eqnarray}
\delta R_{cdab} &=& \frac{1}{4}\tilde{\varepsilon}_{cd}\tilde{\varepsilon}_{ab}
\tilde{\varepsilon}^{ef}\tilde{\varepsilon}^{gh}\delta R_{efgh},
\nonumber\\
\delta R_{cdAB} &=& -\frac{1}{4}\tilde{\varepsilon}_{cd}\hat{\varepsilon}_{AB}
\tilde{\varepsilon}^{ab}\hat{\varepsilon}^{CD}\delta R_{abCD},
\nonumber\\
\delta R_{cDab} &=& -\frac{1}{2}\tilde{\varepsilon}_{ab}\tilde{\varepsilon}^{ef}
\delta R_{cDef},
\nonumber
\end{eqnarray}
etc., a lengthy but straightforward calculation yields
\begin{subequations}
\label{Eq:LinCurv}
\begin{eqnarray}
\delta R_{abcd} &=& -\frac{1}{2}\tilde{\varepsilon}_{ab}\tilde{\varepsilon}_{cd}\left[ 
\tilde{\nabla}^e\tilde{\nabla}^f H_{ef} + \tilde{\Box} H + \tilde{k} H \right],
\label{Eq:LinCurv1}\\
\delta R_{aBcd} &=& -\frac{1}{2}\tilde{\varepsilon}_{cd}\left\{
\hat{\nabla}_B 
\left[\tilde{\varepsilon}^{ef} r\tilde{\nabla}_e\left( \frac{1}{r} H_{af} \right) \right]
 - \hat{\varepsilon}_B{}^C\hat{\nabla}_C\left[ \frac{1}{r}\tilde{\nabla}_a(r {\cal F}) 
 + \frac{2}{r}\tilde{\varepsilon}^{ef} r_{ae} h_f \right] \right\},
\label{Eq:LinCurv2}\\ 
\delta R_{abCD} &=&  -\frac{1}{2}\tilde{\varepsilon}_{ab}\hat{\varepsilon}_{CD}
\hat{\Delta} {\cal F},
\label{Eq:LinCurv3}\\
\delta R_{aBcD} &=& -\frac{1}{2}\hat{\nabla}_B\hat{\nabla}_D H_{ac}
 + \frac{1}{2}\hat{g}_{BD}\left[ r r^b\left( 
 \tilde{\nabla}_a H_{cb} + \tilde{\nabla}_c H_{ab} - \tilde{\nabla}_b H_{ac} \right)
  - \frac{r}{2}\tilde{\nabla}_a\tilde{\nabla}_c(r J) - \frac{r}{2} r_{ac} J \right]
\nonumber\\
 &+& \hat{\varepsilon}_{(B}{}^E\hat{\nabla}_{D)}\hat{\nabla}_E
   \left( \tilde{\nabla}_{(a} h_{c)} \right)
 - \frac{1}{4}\tilde{\varepsilon}_{ac}\hat{\varepsilon}_{BD}\hat{\Delta}{\cal F}, 
\label{Eq:LinCurv4}\\
\delta R_{aBCD} &=& \frac{1}{2}\hat{\varepsilon}_{CD}\left\{
 \frac{1}{2}\hat{\varepsilon}_B{}^E\hat{\nabla}_E
 \left[ r^2\tilde{\nabla}_a J - 2 r r^b H_{ab} \right]
- \hat{\nabla}_B\left[ r r^b\tilde{\varepsilon}_{ab}{\cal F} + \hat{\Delta} h_a
 + 2N h_a \right] \right\},
\label{Eq:LinCurv5}\\
\delta R_{ABCD} &=& r^2\hat{\varepsilon}_{AB}\hat{\varepsilon}_{CD}\left[
r^a r^b H_{ab} - \frac{1}{4}\hat{\Delta} J - \frac{1}{2} r r^a\tilde{\nabla}_a J - N J
 + \frac{1}{2} J \right],
\label{Eq:LinCurv6}
\end{eqnarray}
\end{subequations}
where $H:=\tilde{g}^{ab} H_{ab}$ denotes the trace of $H_{ab}$ and where we have defined
\begin{equation}
{\cal F} := r^2\tilde{\varepsilon}^{ab}\tilde{\nabla}_a\left( \frac{h_b}{r^2} \right).
\end{equation}
The remaining components of the variation of the curvature tensor are obtained from Eqs.~(\ref{Eq:LinCurv}) using the symmetry $\delta R_{\alpha\beta\mu\nu} = \delta R_{[\alpha\beta][\mu\nu]} = \delta R_{\mu\nu\alpha\beta}$.

The linearized Ricci tensor follows from Eqs.~(\ref{Eq:LinCurv}) using the formula
\begin{equation*}
\delta R_{\alpha\beta} = \delta (g^{\mu\nu} R_{\mu\alpha\nu\beta} )
 = g^{\mu\nu}\delta R_{\mu\alpha\nu\beta} 
 - R_{\mu\alpha\nu\beta} g^{\mu\sigma} g^{\nu\tau}\delta g_{\sigma\tau}.
\end{equation*}
Explicitly, this yields
\begin{subequations}
\label{Eq:LinRicci}
\begin{eqnarray}
\delta R_{ab} &=& \frac{r^c}{r}\left( \tilde{\nabla}_a H_{bc}
 + \tilde{\nabla}_b H_{ac} - \tilde{\nabla}_c H_{ab} \right)
 - \frac{1}{2r^2}\,\hat{\Delta} H_{ab} + \tilde{k} H_{ab}
 + \frac{1}{2}\,\tilde{g}_{ab}
   \left( \tilde{\nabla}^c\tilde{\nabla}^d H_{cd} + \tilde{\Box} H - \tilde{k} H\right)
\nonumber\\
&-& \frac{1}{2r^2}\,\tilde{\nabla}_{(a} \left( r^2\tilde{\nabla}_{b)} J \right),
\label{Eq:LinRicci1}\\
\delta R_{aB} &=& \frac{1}{2}\hat{\nabla}_B\left[
  \tilde{\nabla}^b H_{ab} - r\tilde{\nabla}_a\left( \frac{H}{r} \right) 
  - \frac{1}{2}\tilde{\nabla}_a J \right]
  - \frac{1}{2}\hat{\varepsilon}_B{}^C\hat{\nabla}_C\left[
  \frac{1}{r^2}\tilde{\varepsilon}_a{}^b\tilde{\nabla}_b(r^2{\cal F})
  + \frac{\tilde{\Delta}(r^2)}{r^2} h_a + \frac{\hat{\Delta} h_a}{r^2} \right],
\label{Eq:LinRicci2}\\
\delta R_{AB} &=& -\frac{1}{2}(\hat{\nabla}_A\hat{\nabla}_B H)^{TF}
 + \hat{g}_{AB}\left[ \tilde{\nabla}^a(r r^b H_{ab}) - \frac{1}{2}\, r r^a\tilde{\nabla}_a H 
 - \frac{1}{4}\hat{\Delta}(H + J) + \frac{1}{4}\tilde{\Box}(r^2 J) \right]
 \nonumber\\
 &+& \hat{\varepsilon}_{(A}{}^C\hat{\nabla}_{B)}\hat{\nabla}_C (\tilde{\nabla}^a h_a).
\label{Eq:LinRicci3}
\end{eqnarray}
\end{subequations}

Finally, we compute the variation of the Einstein tensor using
\begin{equation*}
\delta G_{\alpha\beta} 
 = \delta R_{\alpha\beta} - \frac{1}{2} g_{\alpha\beta} g^{\mu\nu}\delta R_{\mu\nu}
 + \frac{1}{2} g_{\alpha\beta} G^{\mu\nu}\delta g_{\mu\nu}
 + \frac{1}{2} G
  \left( \delta g_{\alpha\beta} 
   - \frac{1}{2} g_{\alpha\beta} g^{\mu\nu}\delta g_{\mu\nu} \right),
\end{equation*}
where here, $G := g^{\mu\nu} G_{\mu\nu}$ denotes the trace of the Einstein tensor. From this, we first find
\begin{eqnarray*}
\delta G_{ab} &=& (\delta R_{ab})^{tf} + \frac{1}{2} G H_{ab}^{tf}
+ \frac{1}{2}\tilde{g}_{ab}\left[ -\frac{1}{r^2}\hat{g}^{CD}\delta R_{CD} 
 + G^{cd} H_{cd}^{tf} + \frac{1}{2}\tilde{g}^{cd} G_{cd} (H-J) \right],\\
\delta G_{aB} &=& \delta R_{aB} 
+ \frac{1}{2} G \hat{\varepsilon}_B{}^C\hat{\nabla}_C h_a,\\
\delta G_{AB} &=& (\delta R_{AB})^{TF} + \frac{r^2}{2}\hat{g}_{AB}\left[
 -\tilde{g}^{cd}\delta R_{cd} + G^{cd} H^{tf}_{cd} 
 - \frac{1}{2r^2}\hat{g}^{CD} G_{CD} (H-J) \right].
\end{eqnarray*}
Introducing into this the expressions~(\ref{Eq:LinRicci}) for the linearized Ricci tensor and combining the result with the expressions~(\ref{Eq:LinTmunu}) for the linearized stress-energy tensor and the background equations~(\ref{Eq:EinsteinSph}), the linearized Einstein equations, $\delta G_{\mu\nu} = \kappa \delta T_{\mu\nu}$ yield the following set of equations on the two-manifold $\tilde{M}$,
\begin{subequations}
\label{Eq:LinEinsteinOdd}
\begin{eqnarray}
-\tilde{\varepsilon}_a{}^b\tilde{\nabla}_b( r^2{\cal F} ) - (\hat{\Delta} + 2) h_a
 &=& 2\kappa r^2(\nu_a - P h_a),
\label{Eq:LinEinsteinOdd1}\\
\tilde{\nabla}^a h_a &=& 2\kappa r^2\beta,
\label{Eq:LinEinsteinOdd2}
\end{eqnarray}
\end{subequations}
and
\begin{subequations}
\label{Eq:LinEinsteinEven}
\begin{eqnarray}
\left[ 2r\tilde{\nabla}_{(a} (r^c H_{b)c}^{tf} ) - r r^c\tilde{\nabla}_c H_{ab}^{tf}
-\frac{1}{2}\hat{\Delta} H_{ab}^{tf} + r r_{(a}\tilde{\nabla}_{b)} H
- \frac{1}{2}\,\tilde{\nabla}_{(a} \left( r^2\tilde{\nabla}_{b)} J \right) \right]^{tf}
 &=& \kappa r^2\left[ \tau_{ab} - T_{(a}{}^c H_{b)c}^{tf} \right]^{tf},
\label{Eq:LinEinsteinEven1}\\
\tilde{\nabla}^b H_{ab}^{tf} - \frac{r^2}{2}\tilde{\nabla}_a\left( \frac{H}{r^2} \right) 
 - \frac{1}{2}\tilde{\nabla}_a J &=& 2\kappa\mu_a,
\label{Eq:LinEinsteinEven2}\\
-2\tilde{\nabla}^a( r r^b H_{ab}^{tf} ) 
+ \frac{1}{2}(\hat{\Delta} - 2) H + \frac{1}{2}(\hat{\Delta} + 2) J
+ \frac{1}{2r^2}\tilde{\nabla}^a (r^4\tilde{\nabla}_a J)
 &=& \kappa r^2\left( \tilde{g}^{ab}\tau_{ab} - T^{ab} H_{ab}^{tf} \right),
\label{Eq:LinEinsteinEven3}\\
\frac{2}{r}\tilde{\nabla}^a(r^b H_{ab}^{tf} ) + \tilde{\nabla}^a\tilde{\nabla}^b H_{ab}^{tf}
 + \frac{1}{2}\tilde{\Box} H - \frac{1}{2r^2}\hat{\Delta} H
 - \frac{1}{2r^2}\tilde{\nabla}^a( r^2\tilde{\nabla}_a J )
 &=& -\kappa(\lambda - PJ + PH),
\label{Eq:LinEinsteinEven4}\\
 H &=& -4\kappa r^2\alpha.
\label{Eq:LinEinsteinEven5} 
\end{eqnarray}
\end{subequations}
By replacing $h_a$ with $h_a^{(inv)}$, $\nu_a$ with $\nu_a^{(inv)}$ etc. we obtain the corresponding equations in gauge-invariant form, as discussed at the end of the previous subsection. For notational simplicity, we omit the superscript $^{(inv)}$ in what follows.

We see that the linearized Einstein equations decouple into two groups: the first group comprises Eqs.~(\ref{Eq:LinEinsteinOdd}) for the quantities $h_a$, $\nu_a$ and $\beta$ and describes perturbations with {\em odd parity}, sometimes also called {\em axial} perturbations. The second group comprises Eqs.~(\ref{Eq:LinEinsteinEven}) for the remaining perturbations amplitudes and describes {\em even-parity} perturbations, sometimes also called {\em polar} perturbations.

%%%%%%%%%%%%%%%%%%%%%%%%%%%%%%%%%%%%%%%%%%%%%
\subsection{The linearized divergence law}
\label{Sec:Bianchi}

Before we proceed with analyzing the equations, we work out the variation of the divergence law $\nabla^{\mu} T_{\mu\nu} = 0$ for the stress-energy tensor. Using
\begin{equation*}
\delta(\nabla_\mu T^{\mu}{}_\beta) = \partial_\mu \delta T^{\mu}{}_\beta
 + \Gamma^\mu{}_{\mu\alpha}\delta T^\alpha{}_\beta
 - \Gamma^{\alpha}{}_{\mu\beta}\delta T^\mu{}_{\alpha}
 + T^\alpha{}_\beta\delta\Gamma^\mu{}_{\mu\alpha}
 - T^\mu{}_{\alpha}\delta\Gamma^{\alpha}{}_{\mu\beta},
\end{equation*}
the formula $\delta T^\alpha{}_\beta = \delta (g^{\alpha\mu} T_{\mu\beta} )
 = g^{\alpha\mu}\delta T_{\mu\beta} - g^{\alpha\nu} T^{\mu}{}_{\beta}\delta g_{\nu\mu}$, and the expressions~(\ref{Eq:LinTmunu}) we find the following equations:
\begin{equation}
\frac{1}{r^2}\tilde{\nabla}^a\left[ r^2(\nu_a - P h_a) \right] + (\hat{\Delta} + 2)\beta = 0,
\label{Eq:LinDivergenceLawOdd}
\end{equation}
in the odd-parity sector and
\begin{subequations}
\label{Eq:LinDivergenceLawEven}
\begin{eqnarray}
&& \frac{1}{r^2}\tilde{\nabla}^a( r^2\mu_a ) + (\hat{\Delta} + 2)\alpha 
 + \frac{1}{2}(\lambda - P J + P H) - \frac{1}{2} T^{ab} H_{ab} = 0,
\label{Eq:LinDivergenceLawEven1}\\
&& \frac{1}{r^2}\tilde{\nabla}^a
  \left[ r^2(\tau_{ab} - H_{ac}^{tf} T^c{}_b) \right] - \frac{r_b}{r}(\lambda - P J + P H)
 + \frac{1}{r^2}\hat{\Delta}\mu_b 
 - \frac{1}{2} T^{cd}\tilde{\nabla}_b H_{cd}
 + \frac{1}{2}( T^a{}_b - P\delta^a{}_b )\tilde{\nabla}_a J = 0,
\label{Eq:LinDivergenceLawEven2}
\end{eqnarray}
\end{subequations}
in the even-parity sector.

As a consequence of the twice contracted Bianchi identities, $\nabla^\mu G_{\mu\nu} = 0$, similar equations hold for the linearized Einstein tensor. Therefore, the Eqs.~(\ref{Eq:LinDivergenceLawOdd},\ref{Eq:LinDivergenceLawEven}) show that the linearized Einstein equations~(\ref{Eq:LinEinsteinOdd},\ref{Eq:LinEinsteinEven}) are not independent from each other. For example, Eq.~(\ref{Eq:LinDivergenceLawOdd}) shows that taking the divergence on both sides of Eq.~(\ref{Eq:LinEinsteinOdd1}) gives an equivalent equation than the one obtained by applying the operator $\hat{\Delta} + 2$ on both sides of Eq.~(\ref{Eq:LinEinsteinOdd2}). Since $\hat{\Delta} + 2$ is invertible on the space of mono-dipole-free perturbations, we conclude that it is sufficient to impose Eq.~(\ref{Eq:LinEinsteinOdd1}) in the odd-parity sector. Similarly, we see that it is sufficient to impose Eqs. (\ref{Eq:LinEinsteinEven1},\ref{Eq:LinEinsteinEven2},\ref{Eq:LinEinsteinEven3}) in the even-parity sector, the remaining Eqs. (\ref{Eq:LinEinsteinEven4},\ref{Eq:LinEinsteinEven5}) are consequences of the former and the linearized divergence law, Eqs.~(\ref{Eq:LinDivergenceLawEven}) above.

%%%%%%%%%%%%%%%%%%%%%%%%%%%%%%%%%%%%%%%%%%%%%
\section{Vacuum perturbations}
\label{Sec:Vac}
%%%%%%%%%%%%%%%%%%%%%%%%%%%%%%%%%%%%%%%%%%%%%

In this section, we analyze the linear perturbations in vacuum. By Birkhoff's theorem (see, for instance, \cite{Straumann-Book}), the background solution must be the Schwarzschild spacetime. The following derivation of the Regge-Wheeler and Zerilli equations does not assume particular coordinates on $\tilde{M}$ or $S^2$.

%%%%%%%%%%%%%%%%%%%%%%%%%%%%%%%%%%%%%%%%%%%%%
\subsection{Vacuum perturbations with odd parity: The Regge-Wheeler master equation}
\label{SubSec:VacOddRW}

Linear perturbations with odd parity in vacuum are described by Eqs.~(\ref{Eq:LinEinsteinOdd}) with $\nu_a=0$, $\beta=0$ and $P=0$.
For the following, it is convenient to introduce the coordinate-free notation of differential forms. For a one-form ${\bm\omega} = \omega_a dx^a$ on $\tilde{M}$, we have
\begin{equation*}
\tilde{*}{\bm\omega} = -\tilde{\epsilon}_a{}^b\omega_b dx^a,\qquad
\tilde{*}d{\bm\omega} = \tilde{\epsilon}^{ab}\tilde{\nabla}_a\omega_b,\qquad
\tilde{d}^\dagger{\bm\omega} = -\tilde{\nabla}^a\omega_a,
\end{equation*}
where $\tilde{*}$ and $\tilde{d}^\dagger = \tilde{*} d \tilde{*}$ denote, respectively, the Hodge dual and the co-differential operator on $\tilde{M}$, respectively. With this notation, Eq.~(\ref{Eq:LinEinsteinOdd1}) reads
\begin{equation}
\tilde{*} d(r^2 {\cal F}) - (\hat{\Delta} + 2) {\bf h} = 0,
\label{Eq:LinEinsteinOdd1Bis}
\end{equation}
where we recall that ${\cal F} := r^2\tilde{\varepsilon}^{ab}\tilde{\nabla}_a( r^{-2} h_b ) = r^2\tilde{*} d(r^{-2} {\bf h})$ and where ${\bf h}$ denotes the one-form ${\bf h}:=h_a dx^a$. Applying the operator $\tilde{*} d r^{-2}$ on both sides of this equation gives
\begin{equation*}
\tilde{d}^\dagger\left[ \frac{1}{r^2} d(r^2{\cal F}) \right] 
 - \frac{1}{r^2}(\hat{\Delta} + 2){\cal F} = 0.
\end{equation*}
Setting $\Phi := r{\cal F}$ and using the background equation $\tilde{\Delta} r = (1 - N)/r = 2m/r^2$ yields the following scalar equation on $\tilde{M}$,
\begin{equation}
\tilde{\Box}\Phi + \frac{1}{r^2}\left[ -\hat{\Delta} - \frac{6m}{r} \right]\Phi = 0,
\label{Eq:ReggeWheeler}
\end{equation}
where $\tilde{\Box} = \tilde{d}^\dagger d$ is the covariant d'Alembertian on $(\tilde{M},\tilde{{\bf g}})$ and $m$ the Schwarzschild mass. Eq.~(\ref{Eq:ReggeWheeler}) is the covariant form of the Regge-Wheeler equation. Once a solution for $\Phi$ is known, the metric perturbation ${\bf h}$ may be reconstructed using Eq.~(\ref{Eq:LinEinsteinOdd1Bis}) since the operator $\hat{\Delta} + 2$ is invertible on the space of mono-dipole-free perturbations. The one-form ${\bf h}$ obtained in this way automatically satisfies Eq.~(\ref{Eq:LinEinsteinOdd2}), since Eq.~(\ref{Eq:LinEinsteinOdd1Bis}) implies that $\tilde{d}^\dagger{\bf h} = 0$.

%%%%%%%%%%%%%%%%%%%%%%%%%%%%%%%%%%%%%%%%%%%%%
\subsection{Vacuum perturbations with even parity: The Regge-Wheeler and Zerilli master equations}
\label{SubSec:VacEvenRWZ}

In the even-parity sector, Eq.~(\ref{Eq:LinEinsteinEven5}) in vacuum implies that $H=0$, and therefore, $H_{ab} = H_{ab}^{tf}$ is symmetric and trace-free. For the following, it is useful to introduce the one-form
\begin{equation}
{\bf C} := H_{ab} r^a dx^b.
\label{Eq:CDef}
\end{equation}
As long as $N = r^a r_a\neq 0$, this one-form contains the same information as $H_{ab}$ since
\begin{equation*}
H_{ab} = \frac{1}{N}\left( r_a C_b + r_b C_a - \tilde{g}_{ab} r^c C_c \right)
\end{equation*}
if $H_{ab}$ is trace-free. Next, consider Eq.~(\ref{Eq:LinEinsteinEven2}). We may contract this equation once with $r^a$ and once with $\tilde{\varepsilon}^{ab} r_b$. The result is, in coordinate-independent notation,
\begin{eqnarray*}
\tilde{d}^\dagger {\bf C} + \frac{1}{2}\tilde{{\bf g}}(dr,dJ) &=& 0,\\
d{\bf C} - \frac{1}{2} dr \wedge dJ &=& 0,
\end{eqnarray*}
respectively, where we have used the background equation $(r_{ab})^{tf} = 0$. The second of these equations motivates the definition of the following one-form,
\begin{equation}
{\bf Z} := {\bf C} - \frac{r}{2} dJ,
\end{equation}
in terms of which the above equations read $2\tilde{d}^\dagger {\bf Z} - r\tilde{\Box} J = 0$ and $d{\bf Z} = 0$, respectively. Next, Eq.~(\ref{Eq:LinEinsteinEven3}) gives
\begin{equation}
4\tilde{d}^\dagger (r {\bf Z}) + r^2\tilde{\Box} J + (\hat{\Delta} + 2) J = 0.
\label{Eq:LinEvenVacPrep}
\end{equation}
Finally, we contract Eq.~(\ref{Eq:LinEinsteinEven1}) with $r^a$ and use the background equations $2r r_{ab} = \tilde{g}_{ab} (1-N)$ and Eq.~(\ref{Eq:LinEvenVacPrep}) in order to eliminate $\tilde{\Box} J$ from the resulting equation. This yields Eq.~(\ref{Eq:LinEvenVac4}) below.

In conclusion, we obtain the following set of equations governing even-parity linear perturbations of the Schwarzschild solution,
\begin{subequations}
\label{Eq:LinEvenVac}
\begin{eqnarray}
&& d{\bf Z} = 0,
\label{Eq:LinEvenVac1}\\
&& 2\tilde{d}^\dagger {\bf Z} + r\tilde{\Box} J = 0,
\label{Eq:LinEvenVac2}\\
&& 4\tilde{d}^\dagger( r {\bf Z}) + r^2\tilde{\Box} J + (\hat{\Delta} + 2) J = 0,
\label{Eq:LinEvenVac3}\\
&& d\left[ 2r\tilde{g}(dr,{\bf Z}) + 3m J - \frac{1}{2}(\hat{\Delta} + 2)(rJ) \right] - \hat{\Delta} {\bf Z} 
 = 0. \label{Eq:LinEvenVac4}
\end{eqnarray}
\end{subequations}
These equations can be simplified as follows: first, we eliminate $\tilde{\Box} J$ from the second and third equation, giving
\begin{equation}
r^2\tilde{d}^\dagger {\bf Z} - 2r\tilde{g}(dr,{\bf Z}) + \frac{1}{2}(\hat{\Delta} + 2)(rJ) = 0.
\label{Eq:LinEvenVac5}
\end{equation}
Using this, we may rewrite Eq.~(\ref{Eq:LinEvenVac4}) as
\begin{equation}
d[ r^2\tilde{d}^\dagger {\bf Z} + 3m J ] - \hat{\Delta} {\bf Z} = 0.
\label{Eq:LinEvenVac6}
\end{equation}
Now there are two ways to proceed.
\begin{enumerate}
\item We apply the co-differential $\tilde{d}^\dagger$ to Eq.~(\ref{Eq:LinEvenVac6}) and use Eq.~(\ref{Eq:LinEvenVac2}) in order to eliminate $\tilde{\Box} J$ from the resulting equations. This yields the following equation for $\phi:=r^2\tilde{d}^\dagger {\bf Z}$,
\begin{equation}
\tilde{\Box}\phi + \frac{1}{r^2}\left[ -\hat{\Delta} - \frac{6m}{r} \right]\phi = 0,
\end{equation}
which is exactly the covariant form of the Regge-Wheeler equation found in the odd-parity sector.
\item The second way uses the equation $d{\bf Z}=0$ in order to introduce an angular-dependent function $\zeta$ on $\tilde{M}$ such that\footnote{Notice that for the maximal Kruskal extension of the Schwarzschild solution, $\tilde{M}$ is topologically equivalent to $\Real^2$, so this function exists according to Poincar\'e's lemma.} ${\bf Z} = d\zeta$. Then, we may integrate Eq.~(\ref{Eq:LinEvenVac6}) and obtain
\begin{equation}
r^2\tilde{\Box}\zeta + 3m J - \hat{\Delta}\zeta = 0.
\label{Eq:LinEvenVac7}
\end{equation}
Applying the operator $(\hat{\Delta} + 2)$ on both sides of this equation and using 
Eq.~(\ref{Eq:LinEvenVac5}) in order to eliminate $J$ yields
\begin{equation}
\left( \hat{\Delta} + 2 - \frac{6m}{r} \right)\tilde{\Box}\zeta
 + \frac{12m}{r^2}\tilde{{\bf g}}(dr,d\zeta) - \frac{1}{r^2}\hat{\Delta}(\hat{\Delta} + 2)\zeta = 0.
\label{Eq:LinEvenVac8}
\end{equation}
Finally, we define the new scalar $\Psi$ by $\zeta = (\hat{\Delta} + 2 - 6m/r)\Psi$ in order to eliminate the first-order derivatives in Eq.~(\ref{Eq:LinEvenVac8}). This yields the following equation,
\begin{equation}
\tilde{\Box}\Psi + \frac{1}{r^2}\left( \hat{\Delta} + 2 - \frac{6m}{r} \right)^{-2}
\left[ (\hat{\Delta}+2)^2\left( -\hat{\Delta} + \frac{6m}{r} \right)
 + \frac{36m^2}{r^2}\left(-\hat{\Delta} - 2 + \frac{2m}{r} \right) \right]\Psi = 0.
\label{Eq:Zerilli}
\end{equation}
This is the covariant form of the Zerilli equation. It involves the inverse of the linear operator $A(r) := -\hat{\Delta} - 2 + 6m/r$ which is invertible on the mono-dipole-free space. Integrating Eq.~(\ref{Eq:LinEvenVac4}), one finds the following explicit expression for the scalar $\Psi$ in terms of the metric perturbations $H_{ab}$ and $J$:
\begin{equation}
\Psi = \hat{\Delta}^{-1} \left[ \left( \hat{\Delta} + 2 - \frac{6m}{r} \right)^{-1}
\left( 2 r r^a r^b H_{ab} - r^2 r^a\tilde{\nabla}_a J \right) - \frac{r}{2} J \right].
\end{equation}
This is the covariant generalization of the Zerilli-Moncrief function, see~\cite{kMeP05} and references therein.
\end{enumerate}

Although the Zerilli equation is more complicated than the Regge-Wheeler equation, its advantage relies in the fact that the metric perturbations $H_{ab}$ and $J$ can be reconstructed from the scalar $\Psi$ without solving additional differential equations. Indeed, from $\Psi$ we can compute $\zeta = (\hat{\Delta} + 2 - 6m/r)\Psi$, and knowing $\zeta$, we obtain the one-form ${\bf Z} = d\zeta$ and the scalar $J$ from Eq.~(\ref{Eq:LinEvenVac5}). From this, one obtains ${\bf C} = {\bf Z} + r dJ/2$ and then $H_{ab}$. Using Eq.~(\ref{Eq:LinEvenVac8}) in order to eliminate $\tilde{\Box}\zeta$ the explicit result of these operations, first given in Ref.~\cite{lBoS07}, is
\begin{eqnarray*}
J &=& 2\left( \hat{\Delta} + 2 - \frac{6m}{r} \right)^{-1}\left[
 2\tilde{\bf g}(dr,d\zeta) - \frac{1}{r}\hat{\Delta}\zeta \right],\\
H_{ab} &=& 2\left( \hat{\Delta} + 2 - \frac{6m}{r} \right)^{-1}(\tilde{\nabla}_a\tilde{\nabla}_b)^{tf} (r\zeta).
\end{eqnarray*}
In contrast to this, the reconstruction of $H_{ab}$ and $J$ from the scalar $\phi = r^2\tilde{d}^\dagger {\bf Z}$ introduced in method 1 is more involved: Integrating Eq.~(\ref{Eq:LinEvenVac6}) using ${\bf Z} = d\zeta$, one first obtains $3m J = \hat{\Delta}\zeta - \phi$, from which $J$ can be eliminated in Eq.~(\ref{Eq:LinEvenVac5}). This leads to the following differential equation for $\zeta$,
\begin{equation*}
12 m\tilde{{\bf g}}(dr,d\zeta) - (\hat{\Delta}+2)\hat{\Delta}\zeta 
 = \frac{6m}{r}\phi - (\hat{\Delta}+2)\phi,
\end{equation*} 
which could be used to determine the Zerilli potential $\zeta$ and the function $J$ from $\phi$. However, for $m > 0$ the operator on the left-hand side has a non-vanishing kernel, consisting of superpositions of functions of the form
\begin{equation*}
\zeta_{LM}(r,\vartheta,\varphi) 
 = e^{\omega_{L,m} r^*} Y^{LM}(\vartheta,\varphi),
\end{equation*}
where here $\omega_{\ell,m} = (L-1)L(L+1)(L+2)/(12m)$ are the algebraic special frequencies~\cite{sC84}, $r^* = r + 2m\log(r/2m-1)$ the Regge-Wheeler tortoise coordinate and $Y^{LM}$ denote the standard spherical harmonics on $S^2$.

%%%%%%%%%%%%%%%%%%%%%%%%%%%%%%%%%%%%%%%%%%%%%
\section{Metric perturbations coupled to matter fields}
\label{Sec:Mat}
%%%%%%%%%%%%%%%%%%%%%%%%%%%%%%%%%%%%%%%%%%%%%

Although the Zerilli approach described in the previous section is well-suited for describing vacuum perturbations with even parity, it is unclear whether or not it can be generalized to the coupling of matter fields since it is based on the integration of Eq.~(\ref{Eq:LinEvenVac1}) and the introduction of the Zerilli potential $\zeta$. Indeed, when matter fields are present, it might not always be possible to replace the symmetric tensor field $H_{ab}$ by a scalar potential $\zeta$, as is the case in vacuum.\footnote{Important exceptions to this observation include backgrounds which are static, see for example Refs.~\cite{vM74c,vM74d,vM75} and \cite{cMoS03} for the coupling of gravity to linear or non-linear electromagnetic fields.} Here we discuss an alternative approach, which does not rely on the introduction of the Zerilli potential and might be more amendable to the coupling of matter fields. In fact, as we show in Sec.~\ref{Sec:Fluid} below, it naturally leads to a constrained wave system of equations for the case of the linearized Einstein-Euler equations.

The method we discuss here is based on the observation that in Eq.~(\ref{Eq:LinEinsteinEven1}) the gradient of $J$ can be eliminated by means of Eq.~(\ref{Eq:LinEinsteinEven2}) and the trace of ${\bf H}$ can be eliminated taking into account Eq.~(\ref{Eq:LinEinsteinEven5}). This leads to the following wave-like equation for the trace-free part, $\Hz_{ab} := H_{ab}^{tf}$, of the symmetric tensor field $H_{ab}$:
\begin{equation}
{\cal L}_{ab}[\Hbz] = 2\kappa \left[ \tau_{ab}
 - \frac{1}{2}\tilde{g}^{cd} T_{cd}\Hz_{ab}
 - \frac{2}{r^2} \tilde{\nabla}_{(a}\left( r^2 \mu_{b)} \right)
 + 2r^2\tilde{\nabla}_a\tilde{\nabla}_b\alpha + 12r r_{(a}\tilde{\nabla}_{b)}\alpha
 + 8 r_a r_b\alpha \right]^{tf},
\label{Eq:LinEinsteinEvenHab}
\end{equation}
where the linear differential operator ${\cal L}$ is defined as
\begin{equation*}
{\cal L}_{ab}[\Hbz] := -\tilde{\nabla}^c\tilde{\nabla}_c\Hz_{ab}
 + \frac{4}{r}\left( r^c \tilde{\nabla}_{(a}\Hz_{b)c}
 - r_{(a}\tilde{\nabla}^c\Hz_{b)c}
 - \frac{1}{2} r^c \tilde{\nabla}_c\Hz_{ab} \right) 
 + \left( 2\tilde{k} + 2\frac{\tilde{\Delta}r}{r} - \frac{\hat{\Delta}}{r^2} \right)\Hz_{ab}.
\end{equation*}
In deriving this equation, we have used the identity
\begin{equation*}
\left[ 2\tilde{\nabla}_{(a}\left( \tilde{\nabla}^c\Hz_{b)c} \right) \right]^{tf}
= \left( \tilde{\nabla}^c\tilde{\nabla}_c - 2\tilde{k} \right)\Hz_{ab},
\end{equation*}
which is valid for sufficiently smooth symmetric, trace-free tensor fields $\Hz_{ab}$. 

The operator ${\cal L}$ has an interesting symmetry. In order to describe it, we introduce the left-dual of $\Hz_{ab}$, defined as\footnote{One can also consider the right-dual of $\Hz_{ab}$, defined as $(\Hz\tilde{*})_{ab} := \Hz_{ac}\tilde{\varepsilon}^c{}_b$. However, it is easy to prove that $(\Hz\tilde{*})_{ab} = -(\tilde{*}\Hz)_{ab}$, so one does not obtain a fundamentally new transformation.}
\begin{equation*}
(\tilde{*}\Hz)_{ab} := \tilde{\varepsilon}_a{}^c\Hz_{cb}.
\end{equation*}
This dual operator maps the space of symmetric, trace-less tensor fields onto itself, and it is invertible since $(\tilde{*}\tilde{*}\Hz)_{ab} = \Hz_{ab}$. Now, coming back to the symmetry of ${\cal L}$, it can be shown that this operator is invariant with respect to $\tilde{*}$, that is, ${\cal L}_{ab}[\tilde{*}\Hbz] = (\tilde{*}{\cal L})_{ab}[\Hbz]$ for all sufficiently smooth, symmetric, trace-less tensor fields $\Hz_{ab}$. Therefore, it is convenient to expand $\Hz_{ab}$ in a basis which is invariant with respect to $\tilde{*}$, since then the equations for the corresponding coefficients decouple from each other.

Such a basis may be constructed from two future-directed null vectors ${\bf k}$ and ${\bf l}$ with the relative normalization $k^a l_a = -2$. With respect to a suitable orientation, the metric tensor $\tilde{g}_{ab}$ and the volume form $\tilde{\varepsilon}_{ab}$ on $\tilde{M}$ have the form
\begin{equation*}
\tilde{g}_{ab} = -k_{(a} l_{b)},\qquad
\tilde{\varepsilon}_{ab} = + k_{[a} l_{b]},
\end{equation*}
and it follows that $(\tilde{*} k)_a := k_b\tilde{\varepsilon}^b{}_a = k_a$, $(\tilde{*} l)_a := l_b\tilde{\varepsilon}^b{}_a = -l_a$. Therefore, ${\bf k}$ and ${\bf l}$ are eigenvectors of $\tilde{*}$ with eigenvalues $+1$ and $-1$, respectively, and we see that $k_a k_b$ and $l_a l_b$ form a basis for the symmetric, trace-free tensor fields on $\tilde{M}$ which are invariant with respect to the left-dual. As a consequence, we can expand
\begin{equation}
\Hz_{ab} = C k_a k_b + D l_a l_b,\qquad
(\tilde{*}\Hz)_{ab} = -C k_a k_b + D l_a l_b.
\label{Eq:HExpansion}
\end{equation}
The null vectors ${\bf k}$ and ${\bf l}$ are not unique, since they can be rescaled by a function $\chi$ on $\tilde{M}$ according to
\begin{equation}
{\bf k}\mapsto e^\chi {\bf k},\qquad
{\bf l}\mapsto e^{-\chi} {\bf l},
\label{Eq:LocalBoosts}
\end{equation}
corresponding to a local boost with hyperbolic angle $\chi$. Following the GHP formalism~\cite{rGaHrP73} we say that a quantity $f$ has boost-weight $q$ if it transforms like $f\mapsto e^{q\chi} f$ under (\ref{Eq:LocalBoosts}). For instance, in Eq.~(\ref{Eq:HExpansion}), $C$ and $D$ have boost-weight $-2$ and $2$, respectively.

In order to compute covariant derivatives of $\Hz_{ab}$ we first note that there exists a one-form ${\bm\eta} = \eta_a dx^a$ such that
\begin{equation*}
\tilde{\nabla}_a k_b = \eta_a k_b,\qquad 
\tilde{\nabla}_a l_b = -\eta_a l_b,
\end{equation*}
due to the fact that ${\bf k}$ and ${\bf l}$ are null, and due to their relative normalization $k^a l_a = -2 = const.$  With respect to the local boosts, Eq.~(\ref{Eq:LocalBoosts}), the one-form ${\bm\eta}$ transforms like a gauge potential: ${\bm\eta} \mapsto {\bm\eta} + d\chi$. Its invariant part, $d{\bm\eta}$, determines the Gauss curvature of $(\tilde{M},\tilde{\bf g})$:  $\tilde{k} = -\tilde{*} d{\bm\eta} = -\tilde{\varepsilon}^{ab}\tilde{\nabla}_a\eta_b$. The following "creation" and "annihilation" operators
\begin{eqnarray*}
a_q^+ &:=& k^a(\tilde{\nabla}_a - q\eta_a),\\
a_q^- &:=&  l^a(\tilde{\nabla}_a - q\eta_a),
\end{eqnarray*}
map a quantity of boost-weight $q$ to one of boost-weight $q+1$ and $q-1$, respectively. Furthermore, these operators satisfy the commutation relation
\begin{equation}
a_{q-1}^+ a_q^- - a_{q+1}^- a_q^+ = 2q\tilde{k}.
\label{Eq:CommRel}
\end{equation}

Now we are ready to compute the covariant derivatives of $H_{ab}$. Applying $\tilde{\nabla}_c$ on both sides of Eq.~(\ref{Eq:HExpansion}) we find, for instance,
\begin{equation*}
\tilde{\nabla}_c\Hz_{ab} = -\frac{1}{2}\left[
(a_{-2}^-C) k_c k_a k_b + (a_{-2}^+ C) l_c k_a k_b
+ (a_2^- D) k_c l_a l_b + (a_2^+ D) l_c l_a l_b \right].
\end{equation*}
This immediately yields $\tilde{\nabla}^a\Hz_{ab} = (a_{-2}^+ C) k_b + (a_2^- D) l_b$, and using the commutation relation (\ref{Eq:CommRel}) we also find
\begin{equation*}
(-\tilde{\nabla}^c\tilde{\nabla}_c + 2\tilde{k})\Hz_{ab}
 = (a_{-1}^- a_{-2}^+ C) k_a k_b + (a_1^+ a_2^- D) l_a l_b.
\end{equation*}
Using this and expanding $-2r_a = r_{\bf l} k_a + r_{\bf k} l_a$, $r_{\bf k} := k^a r_a$, $r_{\bf l} := l^a r_a$, we find the following decomposition of the operator ${\cal L}$:
\begin{eqnarray}
{\cal L}_{ab}[\Hbz] &=& 
\left[ a_{-1}^- a_{-2}^+ C + 3\frac{r_{\bf l}}{r}(a_{-2}^+ C) - \frac{r_{\bf k}}{r}(a_{-2}^- C)
 + \left( 2\frac{\tilde{\Delta}r}{r} - \frac{\hat{\Delta}}{r^2} \right) C \right] k_a k_b
\nonumber\\
&+& \left[ a_{1}^+ a_{2}^- D + 3\frac{r_{\bf k}}{r}(a_{2}^- D) - \frac{r_{\bf l}}{r}(a_{2}^+ D)
 + \left( 2\frac{\tilde{\Delta}r}{r} - \frac{\hat{\Delta}}{r^2} \right) D \right] l_a l_b.
\label{Eq:Wave-like-eqns_C_D}
\end{eqnarray}
Note the symmetry ${\bf k}\leftrightarrow {\bf l}$ which implies $C\leftrightarrow D$, $a_q^+\leftrightarrow a_{-q}^-$.

%%%%%%%%%%%%%%%%%%%%%%%%%%%%%%%%%%%%%%%%%%%%%
\subsection{Evolution equations (even-parity sector)}
\label{SubSec:Evolution}

The perturbation equation~(\ref{Eq:LinEinsteinEvenHab}) and the decomposition~(\ref{Eq:Wave-like-eqns_C_D}) of the operator ${\cal L}$ yield two wave-like equations for the quantities $C$ and $D$,
\begin{subequations}
\label{Eq:Evol_eqn_CD}
\begin{eqnarray}
a_{-1}^- a_{-2}^+ C + 3\frac{r_{\bf l}}{r}(a_{-2}^+ C) - \frac{r_{\bf k}}{r}(a_{-2}^- C)
 + \left( 2\frac{\tilde{\Delta}r}{r} - \frac{\hat{\Delta}}{r^2} \right) C
 &=& \kappa \Gamma,
 \label{Eq:Evol_eqn_C} \\
a_{1}^+ a_{2}^- D + 3\frac{r_{\bf k}}{r}(a_{2}^- D) - \frac{r_{\bf l}}{r}(a_{2}^+ D)
 + \left( 2\frac{\tilde{\Delta}r}{r} - \frac{\hat{\Delta}}{r^2} \right) D
 &=& \kappa \Delta,
 \label{Eq:Evol_eqn_D}
\end{eqnarray}
\end{subequations}
where
\begin{subequations}
\label{Eq:DefGammaDelta}
\begin{eqnarray}
\Gamma &:=& l^a l^b \left[ \frac{1}{2}\tau_{ab} 
 - \frac{1}{4}\tilde{g}^{cd} T_{cd}\Hz_{ab}
 - \frac{1}{r^2} \tilde{\nabla}_a \left( r^2 \mu_b \right) 
 + r^2\tilde{\nabla}_a\tilde{\nabla}_b\alpha + 6r r_a\tilde{\nabla}_b\alpha
 + 4 r_a r_b\alpha \right],
\\
\Delta &:=& k^a k^b \left[ \frac{1}{2}\tau_{ab} 
 - \frac{1}{4}\tilde{g}^{cd} T_{cd}\Hz_{ab}
 - \frac{1}{r^2} \tilde{\nabla}_a \left( r^2 \mu_b \right) 
 + r^2\tilde{\nabla}_a\tilde{\nabla}_b\alpha + 6r r_a\tilde{\nabla}_b\alpha
 + 4 r_a r_b\alpha \right].
\end{eqnarray}
\end{subequations}

When formulating Eqs.~(\ref{Eq:Evol_eqn_CD}) as a Cauchy problem, one introduces a foliation of $\tilde{M}$ by spacelike hypersurfaces $\Sigma_t$. Let ${\bf u}$ be the future-directed timelike unit normal to $\Sigma_t$, and let ${\bf w}$ be a unitary spacelike vector field orthogonal to ${\bf u}$, which is, therefore, tangent to $\Sigma_t$. We choose the orientation of ${\bf w}$ such that $\tilde{\varepsilon}_{ab} = -2u_{[a} w_{b]}$. Then, we define the null vectors ${\bf k}$ and ${\bf l}$ as
\begin{equation}
{\bf k} := {\bf u} + {\bf w},\qquad 
{\bf  l} := {\bf u} - {\bf w}.
\label{Eq:lkFoliation}
\end{equation}
Notice that from this moment on, we lose the boost-invariance leading to the symmetry described in Eq.~(\ref{Eq:LocalBoosts}), since the foliation singles out a preferred normalization for ${\bf  k}$ and ${\bf l}$. With respect to the $1+1$ decomposition induced by the foliation, the perturbation equations~(\ref{Eq:Evol_eqn_CD}) can be written explicitly as
\begin{subequations}
\label{Eq:CDEvol1+1}
\begin{eqnarray}
\tilde{\Box}C &-& \frac{\hat{\Delta}C}{r^2} - \left( 4\frac{r'}{r} - 2\frac{\dot{r}}{r} - 4\nu \right) \dot{C} + \left( 4\frac{\dot{r}}{r} - 2\frac{r'}{r} - 4\mu \right)C' 
\nonumber\\
&+& \left[ 2\left(  \dot{\nu} - \mu' \right) - 4\left( \mu^2 - \nu^2 \right) + 8 \left( \mu \frac{\dot{r}}{r} - \nu \frac{r'}{r} \right) + 4 \left( \nu \frac{\dot{r}}{r} -\mu \frac{r'}{r} \right) + \frac{8m}{r^3}  \right]C = \kappa\left[ \Gamma + 2(P - \tilde{g}^{ab} T_{ab})C \right],
\label{Eq:CEvol1+1}\\
\tilde{\Box}D &-& \frac{\hat{\Delta}D}{r^2} + \left( 4\frac{r'}{r} + 2\frac{\dot{r}}{r} - 4\nu \right) \dot{D} - \left( 4\frac{\dot{r}}{r} + 2\frac{r'}{r} - 4\mu \right)D' 
\nonumber\\
&+& \left[ 2\left( \mu' - \dot{\nu} \right) - 4\left( \mu^2 - \nu^2 \right) + 8 \left( \mu \frac{\dot{r}}{r} - \nu \frac{r'}{r} \right) + 4 \left( \mu \frac{r'}{r} - \nu \frac{\dot{r}}{r} \right) + \frac{8m}{r^3}  \right]D = \kappa\left[ \Delta + 2(P - \tilde{g}^{ab} T_{ab})D \right].
\label{Eq:DEvol1+1}
\end{eqnarray}
\end{subequations}
Here, a dot and a prime refer to the directional derivatives along ${\bf u}$ and ${\bf w}$, respectively, and $\mu := w^a\eta_a$ and $\nu := u^a\eta_a$. In deriving these equations, we have used the expression $\tilde{k} = -\tilde{\varepsilon}^{ab}\tilde{\nabla}_a\eta_b = \dot{\mu} - \nu' + \mu^2 - \nu^2$ for the Gauss curvature of $(\tilde{M},\tilde{\bf g})$, as well as the background equations $\tilde{k} = \tilde{\Delta} r/r - \kappa P$ and $2\tilde{\Delta} r/r = 4m/r^3 + \kappa\tilde{g}^{ab} T_{ab}$ which follow from Eqs.~(\ref{Eq:EinsteinSph2},\ref{Eq:EinsteinSph3}). Finally, $m$ is the Misner-Sharp mass~\cite{cMdS64} which is defined by $r^a r_a = N = 1 - 2m/r$. Notice the symmetry
\begin{equation}
C\mapsto D,\quad
\mu\mapsto \mu,\quad
\nu\mapsto -\nu,\quad
(\ldots)\dot{ }\mapsto (\ldots)\dot{ },\quad
(\ldots)'\mapsto -(\ldots)',\quad
\Gamma \mapsto \Delta,
\label{Eq:NestorSym}
\end{equation}
which transforms Eq.~(\ref{Eq:CEvol1+1}) to Eq.~(\ref{Eq:DEvol1+1}).
 
Provided that the source terms $\Gamma$ and $\Delta$ do not contain derivatives of $C$ and $D$ of order higher than one, and provided that the quantity $J$ can be eliminated in these terms, Eqs.~(\ref{Eq:CDEvol1+1}) constitute a wave system that, together with suitable evolution equations for the matter variables, allows to determine the evolution of the metric perturbations $C$ and $D$ from initial data for $(C,D)$ and their first time derivatives on an initial time slice.

Notice that the wave equations~(\ref{Eq:CDEvol1+1}) decouple from each other for vacuum perturbations of a Schwarzschild black hole. In this case, an alternative, coordinate-independent way of fixing the boost freedom in the null vectors ${\bf k}$ and ${\bf l}$ is by specifying the two quantities $r_{\bf l}$ and $r_{\bf k}$, which determine the expansion of the two-spheres along in- and outgoing radial null geodesics. These two quantities are constrained by the relation $r_{\bf k} r_{\bf l} = k^{(a} l^{b)} r_a r_b = -\tilde{g}^{ab} r_a r_b = -N = 2m/r - 1$. Note that $r_{\bf k}$ is positive in regions I and III of the Kruskal spacetime and negative in regions II and IV, whereas $r_{\bf l}$ is negative in regions I and II and positive in the remaining ones.\footnote{See, for instance, Fig. 6.9 in Ref.~\cite{Wald-Book} for the Kruskal diagram and the definitions of regions I -- IV.} Therefore, the quantities $r_{\bf k}$ and $r_{\bf l}$ cannot be arbitrarily chosen. The connection coefficients $\eta_{\bf k}$ and $\eta_{\bf l}$ are obtained from $r_{\bf k}$ and $r_{\bf l}$ by solving the background equations~(\ref{Eq:EinsteinSph1},\ref{Eq:EinsteinSph2}) which can be rewritten as $a_1^+(r_{\bf k}) = a_{-1}^-(r_{\bf l}) = 0$, $a_1^-(r_{\bf k}) = a_{-1}^+(r_{\bf l}) = -2m/r^2$.

We choose $r_{\bf k} = N$ and $r_{\bf l} = -1$, which is well-behaved in regions I and II (our universe and the black hole region) of the Kruskal diagram. In terms of ingoing Eddington-Finkelstein coordinates $(v,r)$, for which $\tilde{{\bf g}} = -N dv^2 + 2 dv dr$, the corresponding null vectors are
\begin{equation*}
k^a\frac{\partial}{\partial x^a} 
 = 2\frac{\partial}{\partial v} + N\frac{\partial}{\partial r},\qquad
l^a\frac{\partial}{\partial x^a} = -\frac{\partial}{\partial r},\qquad
k_a dx^a = -N dv + 2dr,\qquad 
l_a dx^a = -dv,
\end{equation*}
and it follows that $\eta_{\bf k} = 2m/r^2$ and $\eta_{\bf l} = 0$. With this choice, Eqs.~(\ref{Eq:Evol_eqn_CD}) reduce to
\begin{eqnarray*}
\tilde{\Box} C  - \frac{\hat{\Delta}}{r^2} C
 - \frac{3}{r} C_{\bf k} - \frac{1}{r}\left( 1 - \frac{6m}{r} \right) C_{\bf l}   &=& 0,\\
\tilde{\Box} D - \frac{\hat{\Delta}}{r^2} D 
 + \frac{1}{r} D_{\bf k} + \frac{1}{r}\left(3 - \frac{10m}{r}\right) D_{\bf l}  &=& 0,
\end{eqnarray*}
with $C_{\bf k} = k^a\tilde{\nabla}_a C$ and $C_{\bf l} = l^a\tilde{\nabla}_a C$.

A different choice which is valid in regions I and III (our universe and the white hole region) of the Kruskal diagram is $r_{\bf k} = \sqrt{2}$, $r_{\bf l }= -N/\sqrt{2}$ which, taking into account changes in the notation and conventions, corresponds to the one made in Ref.~\cite{jBwP73}, where general retarded solutions where computed. With this choice one obtains from Eqs.~(\ref{Eq:Evol_eqn_CD}) their Eqs.~(2.21a,2.21b) with the identifications $C\leftrightarrow r^2\Psi_{-2}$ and $D\leftrightarrow r^2\Psi_2$, where $\Psi_{\pm 2}$ refer to the complex Weyl scalars $\Psi_2 := R_{\alpha\beta\gamma\delta} k^\alpha m^\beta k^\gamma m^\delta$ and $\Psi_{-2} := R_{\alpha\beta\gamma\delta} l^\alpha\bar{m}^\beta l^\gamma\bar{m}^\delta$ computed from a Newman-Penrose null tetrad $\{ {\bf k},{\bf l},{\bf m},\bar{\bf m} \}$.\footnote{In our convention, they are normalized such that ${\bf g}({\bf k},{\bf l}) = -2$ and ${\bf g}({\bf m},{\bf\bar{m}}) = 2$.} Assuming that the background tetrad is adapted to the spherical symmetry in the sense that ${\bf k}$ and ${\bf l}$ are orthogonal to the metric two-spheres, it is relatively simple to compute the first variation of $\Psi_{\pm 2}$  in our perturbation formalism, based on the background expressions~(\ref{Eq:CurvatureBackgr}) and Eq.~(\ref{Eq:LinCurv4}) for the first variation of the mixed components $R_{aBcD}$ of the curvature tensor. The result is
\begin{equation*}
\delta\Psi_2 = -\frac{1}{r^2}\hat{m}^A\hat{m}^B\hat{\nabla}_A\hat{\nabla}_B\left( 
\frac{1}{2} k^a k^b\Hz_{ab} + i k^a k^b\tilde{\nabla}_a h_b \right),
\end{equation*}
where $\hat{\bf m} := r{\bf m}$, and the expression for $\delta\Psi_{-2}$ follows from this after mapping ${\bf k}\mapsto {\bf l}$, ${\bf m}\mapsto \bar{\bf m}$. Therefore, the quantities $k^a k^b\Hz_{ab} = 4D$ and $l^a l^b\Hz_{ab} = 4C$ determine the even-parity contributions of the variation of the complex Weyl scalars $\Psi_2$ and $\Psi_{-2}$, respectively.

%%%%%%%%%%%%%%%%%%%%%%%%%%%%%%%%%%%%%%%%%%%%%
\subsection{Constraint equations (even-parity sector)}
\label{SubSec:Constraints}

The evolution equations in the previous subsection were obtained from the linearized Einstein equation~(\ref{Eq:LinEinsteinEven1}), combined with Eqs.~(\ref{Eq:LinEinsteinEven2}) and (\ref{Eq:LinEinsteinEven5}). Therefore, to solve the whole set of linearized Einstein equations, one has to ensure that the Eqs.~(\ref{Eq:LinEinsteinEven2})~-~(\ref{Eq:LinEinsteinEven5}) are also satisfied. First of all, we impose Eq.~(\ref{Eq:LinEinsteinEven5}) in order to determine the trace $H$ of ${\bf H}$. Next, the Bianchi identities and the zero divergence of the stress-energy tensor imply that Eq.~(\ref{Eq:LinEinsteinEven4}) follows from Eqs.~(\ref{Eq:LinEinsteinEven2},\ref{Eq:LinEinsteinEven5}), as discussed in Sec.~\ref{Sec:Bianchi}. As we now show, the remaining equations, Eqs.~(\ref{Eq:LinEinsteinEven2},\ref{Eq:LinEinsteinEven3}) yield a constraint equation for the evolution system~(\ref{Eq:Evol_eqn_CD}) plus an equation which allows to determine $J$ from the knowledge of $(C,D,\dot{C},\dot{D})$ and the matter fields on a time slice $\Sigma_t$.

In order to derive the constraint equation, we first project Eq.~(\ref{Eq:LinEinsteinEven2}) along the null directions ${\bf k}$ and ${\bf l}$, obtaining
\begin{subequations}
\begin{eqnarray}
a_{-2}^+ C + \frac{1}{4}a_0^- J &=& \kappa l^a(r^2\tilde{\nabla}_a\alpha - \mu_a),
\label{Eq:92_forC}\\
a_2^- D + \frac{1}{4}a_0^+ J &=& \kappa k^a(r^2\tilde{\nabla}_a\alpha - \mu_a).
\label{Eq:92_forD}
\end{eqnarray}
\end{subequations}
Applying the operator $a_{-1}^+$ to both sides of Eq.~(\ref{Eq:92_forC}) and $a_1^-$ to both sides of Eq.~(\ref{Eq:92_forD}), and using the commutation relation~(\ref{Eq:CommRel}) with $q=0$, one obtains the relation
\begin{equation}
a_{-1}^+ a_{-2}^+ C  - a_1^- a_2^- D 
 = -2\kappa\tilde{\varepsilon}^{ab}(\tilde{\nabla}_a\mu_b - 2rr_a\tilde{\nabla}_b\alpha)
 =: 2\kappa G .
\label{Eq:Rel_CD1}
\end{equation}
in which $J$ is eliminated. In order to rewrite this equation as a constraint on the spacelike hypersurfaces $\Sigma_t$ of $\tilde{M}$, the second time derivatives ($\ddot{C}$ and $\ddot{D}$) have to be eliminated in the terms $a_{-1}^- a_{-2}^+ C$ and $a_1^- a_2^- D$. For this, we note
\begin{equation*}
a_{-1}^+ a_{-2}^+ C
 = \left[ a_{-1}^- + 2( \tilde{\nabla}_{\bf w} + \eta_{\bf w}) \right] a_{-2}^+ C
 = a_{-1}^- a_{-2}^+ C + 2( \tilde{\nabla}_{\bf w} + \eta_{\bf w})a_{-2}^+ C,
\end{equation*}
where $\tilde{\nabla}_{\bf w} := w^a\tilde{\nabla}_a$ and $\eta_{\bf w} := w^a \eta_a$, and where we have used the definition of the null vectors ${\bf k}$ and ${\bf l}$ in terms of the vector fields ${\bf u}$ and ${\bf w}$, see Eq.~(\ref{Eq:lkFoliation}), in order to write ${\bf k} = {\bf l} + 2{\bf w}$ in the definition of $a_{-1}^+$. The first term on the right-hand side, $a_{-1}^- a_{-2}^+ C$, can be eliminated using the evolution equation~(\ref{Eq:Evol_eqn_C}). This yields
\begin{equation}
a_{-1}^+ a_{-2}^+ C
= \left[ 2\left( \tilde{\nabla}_{\bf w} + \eta_{\bf w} \right) - \frac{1}{r} (3r_{\bf l} - r_{\bf k}) \right] \left( a_{-2}^+ C \right)
 - 2\frac{r_{\bf k}}{r} \left( \tilde{\nabla}_{\bf w} + 2\eta_{\bf w} \right)C 
 - \left( 2\frac{\tilde{\Delta}r}{r} - \frac{\hat{\Delta}}{r^2} \right) C 
+ \kappa\Gamma,
\label{Eq:aaC}
\end{equation}
where we have also used $a_{-2}^- C = a_{-2}^+ C - 2\left( \tilde{\nabla}_{\bf w} + 2\eta_{\bf w} \right)C$ for later convenience. Similarly, one obtains, using  the evolution equation~(\ref{Eq:Evol_eqn_D}).
\begin{equation}
a_1^- a_2^- D
= -\left[ 2\left( \tilde{\nabla}_{\bf w} - \eta_{\bf w} \right) + \frac{1}{r} (3r_{\bf k} - r_{\bf l}) \right] \left( a_2^- D \right)
 + 2\frac{r_{\bf l}}{r} \left( \tilde{\nabla}_{\bf w} - 2\eta_{\bf w} \right)D 
 - \left( 2\frac{\tilde{\Delta}r}{r} - \frac{\hat{\Delta}}{r^2} \right) D 
+ \kappa\Delta.
\label{Eq:aaD}
\end{equation}
Therefore, provided that the matter terms $\Gamma$, $\Delta$ and $G$ do not contain derivatives of $C$ or $D$ along ${\bf u}$ of order higher than one, Eq.~(\ref{Eq:Rel_CD1}) with the substitutions for $a_{-1}^+ a_{-2}^+ C$ and $a_1^- a_2^- D$ given in Eqs.~(\ref{Eq:aaC},\ref{Eq:aaD}), yields a constraint for the data $(C,D,\dot{C},\dot{D})$ and the matter fields on a given surface $\Sigma_t$.

Next, we consider Eq.~(\ref{Eq:LinEinsteinEven3}), where we use the background equation~(\ref{Eq:EinsteinSph1}), the identity $\tilde{\nabla}^a \tilde{\nabla}^b\Hz_{ab} = a_{-1}^+ a_{-2}^+ C + a_1^- a_2^- D$, as well as Eq.~(\ref{Eq:LinEinsteinEven2}) in order to eliminate the first and second-order derivatives of $J$. This yields the following equation,
\begin{equation}
-(\hat{\Delta} + 2)J = 2r^2( a_{-1}^+ a_{-2}^+ C + a_1^- a_2^- D)
 + 4r( r_{\bf k} a_{-2}^+ C + r_{\bf l} a_2^- D ) - 4(r_{\bf k}^2 C + r_{\bf l}^2 D) - 4r^2\kappa F,
\label{Eq:JEliminate}
\end{equation}
with
\begin{equation}
F := \frac{1}{2}\tilde{g}^{ab}\tau_{ab} - T^{ab}\Hz_{ab} 
 + \tilde{\nabla}^a\mu_a + 4\frac{r^a}{r}\mu_a
 + r^2\tilde{\Box}\alpha + (\hat{\Delta}-2)\alpha - 6r r^a\tilde{\nabla}_a\alpha.
\label{Eq:FDef}
\end{equation}
Provided the derivatives of $J$ can be eliminated in the matter term $F$, this equation allows one to determine $J$ as a function of $(C,D)$ and the matter fields by inverting the operator $-(\hat{\Delta} + 2)$. In fact, using the substitutions for $a_{-1}^+ a_{-2}^+ C$ and $a_1^- a_2^- D$ given in Eqs.~(\ref{Eq:aaC},\ref{Eq:aaD}), it is even possible to compute $J$ from the data $(C,D,\dot{C},\dot{D})$ and the matter fields on a given time slice $\Sigma_t$.

%%%%%%%%%%%%%%%%%%%%%%%%%%%%%%%%%%%%%%%%%%%%%
\subsection{Initial data construction (even-parity sector)}
\label{SubSec:InitialData}

Here, we discuss a practical algorithm for constructing initial data $(C,D,\dot{C},\dot{D})$ for the wave system~(\ref{Eq:Evol_eqn_CD}). We assume for simplicity that the matter terms $\Gamma$, $\Delta$, $F$ and $G$ only depend on $(C,D)$ but not on their derivatives, and that they are completely determined by these quantities and the initial data for the matter fields. Furthermore, we assume that $G$ does not depend on $C$ nor on $D$. For instance, these assumptions are satisfied in the perfect fluid case as will be discussed in the next section.

Our construction specifies $J$ and its time derivative, $\dot{J} := u^a\tilde{\nabla}_a J$, on the initial surface, as well as suitable initial data for the matter fields. As a consequence, the full gradient $\tilde{\nabla}_a J$ is known on $\Sigma_0$, and we can determine the quantities $a_{-2}^+ C$ and $a_2^- D$ from Eqs.~(\ref{Eq:92_forC},\ref{Eq:92_forD}). Next, we combine Eqs.~(\ref{Eq:Rel_CD1},\ref{Eq:aaC},\ref{Eq:aaD},\ref{Eq:JEliminate}) in order to eliminate $a_{-1}^+ a_{-2}^+ C$ and $a_1^- a_2^- D$, and obtain the two equations
\begin{subequations}
\label{Eq:ODECD}
\begin{eqnarray}
&& 2 \frac{r_{\bf k}}{r} \left( \tilde{\nabla}_{\bf w} + 2\eta_{\bf w} \right)C + \left( 2\frac{\tilde{\Delta}r}{r} - \frac{\hat{\Delta}}{r^2} \right) C + \frac{1}{r^2}\left( r_{\bf k}^2 C + r_{\bf l}^2 D \right)
\nonumber\\
 && \qquad =\, \left[ 2\left( \tilde{\nabla}_{\bf w} + \eta_{\bf w} \right) - \frac{1}{r} (3r_{\bf l} - 2r_{\bf k}) \right] \left( a_{-2}^+ C \right)
 + \frac{r_{\bf l}}{r}(a_2^- D) + \frac{1}{4r^2}(\hat{\Delta} + 2)J
 + \kappa(\Gamma - F - G),
\label{Eq:ODECD1}\\
&& 2 \frac{r_{\bf l}}{r} \left( \tilde{\nabla}_{\bf w} - 2\eta_{\bf w} \right)D - \left( 2\frac{\tilde{\Delta}r}{r} - \frac{\hat{\Delta}}{r^2} \right) D - \frac{1}{r^2}\left( r_{\bf k}^2 C + r_{\bf l}^2 D \right)
\nonumber\\
&& \qquad =\, \left[ 2\left( \tilde{\nabla}_{\bf w} - \eta_{\bf w} \right) + \frac{1}{r} (3r_{\bf k} - 2r_{\bf l}) \right] \left( a_2^- D \right)
 - \frac{r_{\bf k}}{r}(a_{-2}^+ C) + \frac{1}{4r^2}(\hat{\Delta} + 2)J
 - \kappa(\Delta - F + G).
\label{Eq:ODECD2}
\end{eqnarray}
\end{subequations}
The first three terms on the right-hand sides of these equations are known, since $a_{-2}^+ C$, $a_2^- D$ and $J$ are known on the initial surface $\Sigma_0$. Furthermore, according to our assumptions, $\Gamma$, $F$ and $G$ depend on $(C,D)$ but not on their derivatives. Therefore, Eqs.~(\ref{Eq:ODECD}) constitute a system of ordinary differential equations for $C$ and $D$. Once this system has been solved, the initial data for $(\dot{C},\dot{D})$ is obtained by setting
\begin{equation}
\dot{C} := -C' - 2\eta_{\bf k} C + a_{-2}^+ C,\qquad
\dot{D} := D' + 2\eta_{\bf l} D + a_2^+ D.
\label{Eq:CdotDdot}
\end{equation}
This provides initial data $(C,D,\dot{C},\dot{D})$ on the initial surface $\Sigma_0$. The evolution equations~(\ref{Eq:CDEvol1+1}), together with suitable evolution equations for the matter fields, determine $(C,D)$ on an arbitrary time slice $\Sigma_t$. The metric field $J$ is obtained from this by solving Eq.~(\ref{Eq:JEliminate}). This determines the metric perturbations in the even-parity sector.

%%%%%%%%%%%%%%%%%%%%%%%%%%%%%%%%%%%%%%%%%%%%%
\subsection{The odd-parity sector}
\label{SubSec:OddParityMatter}

Odd-parity gravitational perturbations are much simpler to describe than even-parity ones, and gauge-invariant master equations for them have been already obtained in the literature, see for instance Refs.~\cite{uGuS79,dBjM09}. Since our approach is more general because it does not require the decomposition into spherical harmonics, and also for completeness, we briefly describe the odd-parity sector here.

In this case, the metric perturbations are determined by the one-form ${\bf h}$ on $\tilde{M}$ whose dynamics is governed by Eqs.~(\ref{Eq:LinEinsteinOdd}). As discussed in Sec.~\ref{Sec:Bianchi}, it is sufficient to consider the equations of motion for the matter fields and Eq.~(\ref{Eq:LinEinsteinOdd1}), which, in terms of the notation of differential forms, reads
\begin{equation}
\tilde{*} d(r^2 {\cal F}) - (\hat{\Delta} + 2) {\bf h} = 2\kappa r^2 {\bf \Upsilon},
\label{Eq:LinEinsteinOddMatter}
\end{equation}
with ${\cal F} = r^2\tilde{*} d(r^{-2} {\bf h})$ and ${\bf \Upsilon} := (\nu_a - P h_a) dx^a$. Now we can proceed exactly as in Sec.~\ref{SubSec:VacOddRW} to derive a master equation for the scalar field $\Phi := r{\cal F}$. Applying the operator $r^4\tilde{*} d r^{-2}$ on both sides of Eq.~(\ref{Eq:LinEinsteinOddMatter}) we first obtain the following inhomogeneous wave-type equation for $r^2{\cal F}$:
\begin{equation}
r^4\tilde{d}^\dagger\left[ \frac{1}{r^2} d(r^2{\cal F}) \right] 
 - (\hat{\Delta} + 2)(r^2{\cal F}) 
 = 2\kappa r^4 \tilde{*} d{\bf \Upsilon},
\label{Eq:LinEinsteinOddMatterF}
\end{equation}
Introducing ${\cal F} = \Phi/r$ and using the background equation~(\ref{Eq:EinsteinSph2}), we then obtain
\begin{equation}
\tilde{\Box}\Phi + \left[ -\frac{\hat{\Delta}}{r^2} - \frac{6m}{r^3} 
 - \frac{\kappa}{2}\tilde{g}^{ab} T_{ab} \right]\Phi 
 = 2\kappa r \tilde{*} d{\bf \Upsilon},
\end{equation}
which generalizes the Regge-Wheeler equation~(\ref{Eq:ReggeWheeler}) to the case of odd-parity linear metric perturbations coupled to matter fields. Provided the one-form ${\bf \Upsilon}$ does not depend on ${\bf h}$, this yields a wave equation for the gauge-invariant scalar $\Phi$ on $\tilde{M}$. Once $\Phi$ is known, ${\bf h}$ can be reconstructed from Eq.~(\ref{Eq:LinEinsteinOddMatter}) on the mono-dipole-free space. In case ${\bf \Upsilon}$ depends on ${\bf h}$, the method just described also works provided Eq.~(\ref{Eq:LinEinsteinOddMatter}) can be solved for ${\bf h}$.

%%%%%%%%%%%%%%%%%%%%%%%%%%%%%%%%%%%%%%%%%%%%%
\section{Fluid perturbations}
\label{Sec:Fluid}
%%%%%%%%%%%%%%%%%%%%%%%%%%%%%%%%%%%%%%%%%%%%%

Here, we apply the ideas described in the previous section to linear perturbations of spherical, self-gravitating fluid configurations. We start in Sec.~\ref{SubSec:BasicFluid} with a short review of the equations of motion for a relativistic perfect fluid in local thermodynamic equilibrium and discuss potential flows. Next, in Sec.~\ref{SubSec:LinFluid} we linearize the flow, assuming that the background is vorticity-free and isentropic, and show that the fluid perturbations can be split into two parts, the first part describing the perturbations of the vorticity and the entropy and the second part describing a linearized potential flow. The resulting equations are specialized to the case of a spherically symmetric background in Sec.~\ref{SubSec:LinFluidSphBack}. In Sec.~\ref{SubSec:LinFluidSET} we linearize the stress-energy tensor. The propagation and constraint equations in the odd- and even-parity sectors are derived in Sec.~\ref{SubSec:LinFluidOdd} and Sec.~\ref{SubSec:LinFluidEven}, respectively.

Previous work regarding the linearization of relativistic perfect fluids include Ref.~\cite{vM80}, where the stability of steady state accretion flows is analyzed and on which most of the ideas presented in Sec.~\ref{SubSec:BasicFluid} and Sec.~\ref{SubSec:LinFluid} are based, Ref.~\cite{kTaC67} which analyze the stability of static stars and Ref.~\cite{cGjM00} which derive a system of propagation equations based on the covariant gauge-invariant approach. Besides the avoidance of the spherical harmonic decomposition, the results presented below differ from the ones obtained in Ref.~\cite{cGjM00} in the strategy for deriving the evolution equations. While in~\cite{cGjM00} projections of the linearized stress-energy tensor which eliminate the matter variables are considered, here we exploit the fact that a spherically symmetric isentropic flow gives rise to a closed one-form ${\bf v}$, which allows to naturally split the fluid perturbations into a linearized potential flow and a complementary part describing the propagation of the linearized vorticity and entropy. Although the results in~\cite{cGjM00} lead to a free (unconstrained) evolution system and do not require the background to be isentropic, we believe that our approach allows for a clearer physical interpretation of the fluid perturbations.

%%%%%%%%%%%%%%%%%%%%%%%%%%%%%%%%%%%%%%%%%%%%%
\subsection{Basic fluid equations and potential flows}
\label{SubSec:BasicFluid}

We consider a perfect, relativistic fluid with isotropic pressure. It is described by the energy density $\epsilon$, the pressure $p$, and the particle density $n$, as measured by an observer co-moving with the fluid elements which have four-velocity ${\bf u}$ (${\bf u}$ is normalized such that $g_{\mu\nu} u^\mu u^\nu = -1$). The equations of motion for the fluid are
\begin{equation}
\nabla_\mu J^\mu = 0,\qquad
\nabla_\mu T^{\mu\nu} = 0,
\label{Eq:FluidEquations}
\end{equation}
where $J^\mu = n u^\mu$ is the particle current density and $T^{\mu\nu} = \epsilon u^\mu u^\nu + p\gamma^{\mu\nu}$, $\gamma^{\mu\nu} := g^{\mu\nu} + u^\mu u^\nu$, denotes the stress-energy tensor. We require local thermodynamic equilibrium, that is, each fluid element is in thermodynamic equilibrium. Therefore, we assume the existence of an equation of state $\epsilon = \epsilon(s,n)$ which satisfies the first law of thermodynamics,
\begin{equation*}
d\left( \frac{\epsilon}{n} \right) = T ds - p d\left( \frac{1}{n} \right),
\end{equation*}
with $T$ the temperature and $s$ the entropy per particle. For the following, it is convenient to replace the energy density $\epsilon$ by the enthalpy $h := (\epsilon + p)/n$ per particle, which satisfies $dh = T ds + dp/n$. With this notation and under these assumptions the fluid equations~(\ref{Eq:FluidEquations}) yield (assuming $T > 0$),
\begin{subequations}
\label{Eq:Fluid}
\begin{eqnarray}
\nabla_{\bf u} s &=& 0\qquad \hbox{(entropy conservation along the fluid lines)},
\label{Eq:Fluid0}\\
\nabla_{\bf u} n + \theta n &=& 0\qquad \hbox{(particle conservation law)},
\label{Eq:Fluid1}\\
h a_\mu + D_\mu h - T D_\mu s &=& 0\qquad \hbox{(relativistic Euler equations)},
\label{Eq:Fluid2}
\end{eqnarray}
\end{subequations}
where $\theta := \nabla_\mu u^\mu$ and $a_\mu := \nabla_{\bf u} u_\mu$ denote the expansion and acceleration of the fluid, respectively, and where $D_\mu h := \gamma_\mu{}^\nu \nabla_\nu h$ denotes the differential of $h$ projected onto the space orthogonal to ${\bf u}$. Equations~(\ref{Eq:Fluid}), together with the equation of state form a closed evolution system for the quantities $s$, $n$ and ${\bf u}$. Eq.~(\ref{Eq:Fluid0}) implies, in particular, that there is no heat transfer between the different fluid elements since $T ds = 0$ along the fluid lines. (However, the entropy $s$ may vary from one fluid trajectory to another.)

In terms of the one-form ${\bf v} := h u_\mu dx^\mu$ and the vorticity $\Omega_{\mu\nu} := \gamma_\mu{}^\alpha\gamma_\nu{}^\beta \nabla_{[\alpha} u_{\beta]}$ of the fluid, the Euler equations~(\ref{Eq:Fluid2}) are equivalent to
\begin{equation}
F_{\mu\nu} := \nabla_\mu v_\nu - \nabla_\nu v_\mu 
 = 2\left( h\Omega_{\mu\nu} - T u_{[\mu} D_{\nu]} s \right).
\label{Eq:FmunuFluid}
\end{equation}
Therefore, the one-form ${\bf v}$ is closed if and only if the flow is irrotational ($\Omega_{\mu\nu} = 0$) and isentropic ($s = const$). In this case, there exists (at least locally) a potential $\psi$ such that ${\bf v} = d\psi$. Then, the particle conservation law, Eq.~(\ref{Eq:Fluid1}), yields
\begin{equation}
\nabla_\mu\left( \frac{n}{h}\nabla^\mu\psi \right) = 0.
\label{Eq:PotentialFlow}
\end{equation}
Here, $n$ is regarded as a function of $h$ obtained by inverting the relation $h = h(n)$ and $h$ is obtained from the condition $v^\mu v_\mu = h^2 u^\mu u_\mu = -h^2$, which yields
\begin{equation}
h = \sqrt{-\nabla^\mu\psi\cdot \nabla_\mu\psi}.
\label{Eq:hDependence}
\end{equation}
For the case of a stiff equation of state, for which $h$ is proportional to $n$, Eq.~(\ref{Eq:PotentialFlow}) reduces to the standard wave equation $\Box_{\bf g}\psi = 0$ on $(M,{\bf g})$ for the potential $\psi$. However, in general Eq.~(\ref{Eq:PotentialFlow}) is a nonlinear wave equation whose characteristics depend on the gradient of $\psi$. Eq.~(\ref{Eq:PotentialFlow}) and the associated stress-energy tensor $T^{\mu\nu} = n h^{-1}(\nabla^\mu\psi)(\nabla^\nu\psi) + p(h) g^{\mu\nu}$ can also be obtained from the simple action functional
\begin{equation}
S[\psi,{\bf g}] := \int p(h) \sqrt{|g|} d^4 x,
\end{equation}
where $h$ is given by Eq.~(\ref{Eq:hDependence}) and $p(h)$ obeys the first law $dp = ndh$.

The linearization of Eq.~(\ref{Eq:PotentialFlow}) yields the remarkably elegant equation~\cite{vM80}
\begin{equation}
\Box_\mathfrak{G}\delta\psi = 0
\label{Eq:LinearizedPotentialFlow}
\end{equation}
for the perturbed potential $\delta\psi$, where $\Box_\mathfrak{G}$ is the wave operator belonging to the {\em sound metric} $\mathfrak{G}$ defined by
\begin{equation}
\mathfrak{G}_{\mu\nu} := \frac{n}{h}\frac{1}{v_s}\left[ g_{\mu\nu} 
 + (1 - v_s^2) u_\mu u_\nu \right],
\qquad
v_s^2 := \frac{\partial p}{\partial\epsilon} = \frac{\partial\log(h)}{\partial\log(n)},
\label{Eq:SoundMetric}
\end{equation}
$v_s$ denoting the sound speed (in units for which $c=1$).\footnote{The following expressions for the inverse sound metric and its determinant are useful:
\begin{equation*}
\mathfrak{G}^{\mu\nu} 
 = \frac{h}{n} v_s\left[ g^{\mu\nu} + \left( 1 - \frac{1}{v_s^2} \right) u^\mu u^\nu \right],
\qquad
\sqrt{|\mathfrak{G}|} = \sqrt{|g|}\left( \frac{n}{h} \right)^2\frac{1}{v_s}.
\end{equation*}}
For the following, we assume that the sound speed satisfies $0 < v_s\leq 1$. By employing an orthonormal tetrad $\{ {\bf e}_0,{\bf e}_1,{\bf e}_2,{\bf e}_3 \}$ with respect to ${\bf g}$ which is adapted to the fluid flow, i.e. such that ${\bf e}_0 = {\bf u}$, it is not difficult to see that $\mathfrak{G}$ is Lorentzian, and that its cone (the sound cone) lies {\em inside} or (if $v_s=1$) {\em coincides with} the light cone of ${\bf g}$. Notice also that ${\bf u}$ is timelike with respect to both ${\bf g}$ and $\mathfrak{G}$.

%%%%%%%%%%%%%%%%%%%%%%%%%%%%%%%%%%%%%%%%%%%%%
\subsection{The linearized fluid equations}
\label{SubSec:LinFluid}

Linearizing Eq.~(\ref{Eq:FmunuFluid}) about an isentropic, irrotational fluid configuration one obtains
\begin{equation}
\nabla_\mu(\delta v_\nu) - \nabla_\nu(\delta v_\mu) 
 = 2\left( h\delta\Omega_{\mu\nu} - T u_{[\mu} D_{\nu]} \delta s \right).
\label{Eq:FmunuFluidPert}
\end{equation}
For the following, it is convenient to decompose $\delta{\bf v}$ into two terms, the first one being the differential of a function $\Psi$ and the second one a one-form ${\bf W}$ which is orthogonal to ${\bf u}$:
\begin{equation}
\delta{\bf v} = d\Psi + {\bf W},\qquad
u^\mu W_\mu = 0.
\label{Eq:vDecomp}
\end{equation}
This decomposition always exists. Indeed, if $\delta{\bf v}$ is given, we may integrate the equation $\pounds_{\bf u}\Psi = u^\mu\delta v_\mu$ along the flow lines to obtain $\Psi$, and set ${\bf W} := \delta{\bf v} - d\Psi$. By construction, $u^\mu W_\mu = 0$ and $\delta{\bf v} = d\Psi + {\bf W}$. The quantities $\Psi$ and ${\bf W}$ are unique up to transformations of the form
\begin{equation}
\Psi \mapsto \Psi + f,\qquad
W_\mu \mapsto W_\mu - D_\mu f,
\label{Eq:FluidGT}
\end{equation}
where $f$ is a function satisfying $\pounds_{\bf u} f = 0$. This freedom can be fixed by restricting the initial data to $\Psi=0$, or by imposing suitable initial conditions on ${\bf W}$.

Introducing the decomposition~(\ref{Eq:vDecomp}) into Eq.~(\ref{Eq:FmunuFluidPert}) and projecting the result orthogonal to ${\bf u}$ gives
\begin{equation}
2h\delta\Omega_{\mu\nu} = D_\mu W_\nu - D_\nu W_\mu,
\label{Eq:VorticityPert}
\end{equation}
which shows that the one-form ${\bf W}$ describes the vorticity part of the perturbations. The remaining information in Eq.~(\ref{Eq:FmunuFluidPert}) is obtained by contracting with $u^\mu$. Using $u^\mu W_\mu = 0$, this yields
\begin{equation}
\pounds_{\bf u} W_\mu = T D_\mu(\delta s),
\label{Eq:WEvol}
\end{equation}
which, together with the linearization of Eq.~(\ref{Eq:Fluid0}),
\begin{equation}
\pounds_{\bf u}(\delta s) = 0,
\label{Eq:sEvol}
\end{equation}
describes the evolution of the variation of the entropy and the one-form ${\bf W}$ along the flow. As a consequence of Eqs.~(\ref{Eq:VorticityPert},\ref{Eq:WEvol}) one also finds
\begin{equation}
\pounds_{\bf u} (h\delta\Omega_{\mu\nu}) 
 = h D_{[\mu}\left( \frac{T}{h} \right) D_{\nu]}\delta s,
\label{Eq:OmegaEvol}
\end{equation}
which shows that in general, a spatial gradient of $\delta s$ generates vorticity along the flow. Eqs.~(\ref{Eq:WEvol}) and (\ref{Eq:sEvol}) describe the evolution of the variations of the entropy and the one-form ${\bf W}$ which describes the variation of the vorticity through the relation~(\ref{Eq:VorticityPert}). These equations form a system of advection equations along the fluid lines, and since they do not depend on $\Psi$ nor on the variation of the metric fields, they decouple from the remaining perturbation equations and can be solved separately.

It remains to linearize Eq.~(\ref{Eq:Fluid1}), which is equivalent to the continuity equation $\nabla_\mu( n h^{-1} g^{\mu\nu} v_\nu) = 0$. The variation can be written in the form
\begin{equation*}
\nabla_\mu \mathfrak{J}^\mu = 0,
\end{equation*}
where the vector field $\mathfrak{J}^\mu$ is given by
\begin{equation}
\mathfrak{J}^\mu = \frac{n}{h}\left[ 
 \frac{h}{n}\delta\left( \frac{n}{h} \right) v^\mu 
  + g^{\mu\nu}\left(\delta v_\nu - v^\beta\delta g_{\nu\beta} \right)
 + \frac{1}{2} v^\mu g^{\alpha\beta}\delta g_{\alpha\beta} \right].
\label{Eq:JVar}
\end{equation}
In order to evaluate the variation of $n/h$, we regard $n = n(h,s)$ as a function of the enthalpy $h$ and the entropy $s$ per particle, which is obtained by formally inverting the relation $h = h(n,s)$. Using the definition of the sound speed $v_s$ and the first law, we find the thermodynamic relations
\begin{equation*}
v_s^2 := \left. \frac{\partial p}{\partial\epsilon} \right|_s
 = \left. \frac{\partial\log h}{\partial\log n} \right|_s,\qquad
\left. \frac{\partial h}{\partial s} \right|_n 
 = T + \frac{1}{n}\left. \frac{\partial p}{\partial s} \right|_n.
\end{equation*}
Using the identities
\begin{equation*}
\left. \frac{\partial p}{\partial s}\right|_v
\left. \frac{\partial s}{\partial v}\right|_p
\left. \frac{\partial v}{\partial p}\right|_s = -1,\qquad
\left. \frac{\partial v}{\partial s}\right|_p =
\left. \frac{\partial v}{\partial T}\right|_p
\left. \frac{\partial T}{\partial s}\right|_p,\qquad v:=\frac{1}{n},
\end{equation*}
and the definitions of the heat capacity per particle at constant pressure $c_p$, the coefficient of thermal expansion $\alpha_p$ and the compressibility at constant entropy $\kappa_s$, defined as (see, for instance, Ref.~\cite{Huang-Book})
\begin{equation*}
c_p := T\left. \frac{\partial s}{\partial T} \right|_p,\qquad
\alpha_p := \frac{1}{v}\left. \frac{\partial v}{\partial T} \right|_p,\qquad
\kappa_s := -\frac{1}{v}\left. \frac{\partial v}{\partial p} \right|_s,
\end{equation*}
the term involving the partial derivative of the pressure can be rewritten as
\begin{equation*}
\left. \frac{\partial p}{\partial s} \right|_n = T\frac{\alpha_p}{c_p\kappa_s}.
\end{equation*}
Gathering the results, we obtain
\begin{equation}
\left. \frac{\partial n}{\partial h} \right|_s = \frac{n}{h}\frac{1}{v_s^2},\qquad
\left. \frac{\partial n}{\partial s} \right|_h 
 = -\frac{n}{h}\frac{1}{v_s^2}\left. \frac{\partial h}{\partial s} \right|_n 
 = -\frac{n}{h}\frac{T}{v_s^2}\Lambda,\qquad
\Lambda := 1 + \frac{1}{n}\frac{\alpha_p}{c_p\kappa_s},
\label{Eq:LambdaDef}
\end{equation}
from which we can finally compute
\begin{equation}
\delta\left( \frac{n}{h} \right) = \frac{n}{h^2}\left[ \left( \frac{1}{v_s^2} - 1 \right)\delta h 
 - \frac{\Lambda T}{v_s^2}\delta s \right].
\label{Eq:nhVar}
\end{equation}
Next, we compute the variation of the identity $h^2 = -g^{\alpha\beta} v_\alpha v_\beta$ which follows from the definition of $v_\mu = h u_\mu$ and the normalization of ${\bf u}$. Taking into account the decomposition~(\ref{Eq:vDecomp}) this gives
\begin{equation}
2h\delta h = -2v^\alpha\delta v_\alpha + v^\alpha v^\beta\delta g_{\alpha\beta}
 = -2v^\alpha\nabla_\alpha\Psi + v^\alpha v^\beta\delta g_{\alpha\beta}.
\label{Eq:hVar}
\end{equation}
Combining Eqs.~(\ref{Eq:JVar},\ref{Eq:nhVar},\ref{Eq:hVar}) and recalling the definition of the sound metric, Eq.~(\ref{Eq:SoundMetric}), we finally arrive at the following expression:
\begin{equation*}
\sqrt{|g|} \mathfrak{J}^\mu
 = \sqrt{|\mathfrak{G}|}\left[ \mathfrak{G}^{\mu\nu}\left( \nabla_\nu\Psi + W_\nu
  - \delta g_{\nu\beta} v^\beta \right) 
  + \frac{1}{2} v^\mu \mathfrak{G}^{\alpha\beta}\delta g_{\alpha\beta}
  - \frac{v^\mu}{n v_s}\Lambda T\delta s \right].
\end{equation*}
Therefore, the linearized continuity equation yields the inhomogeneous wave equation
\begin{equation}
\Box_\mathfrak{G}\Psi = \divrg_\mathfrak{G}(X),
\label{Eq:LinearizedFlow}
\end{equation}
where $\divrg_\mathfrak{G}$ refers to the divergence of a vector field with respect to the Levi-Civita connection associated to the sound metric, and ${\bf X}$ is the vector field defined as
\begin{equation*}
X^\mu = \mathfrak{G}^{\mu\nu}(W_\nu - \delta g_{\nu\beta} v^\beta)
 + \frac{1}{2} v^\mu\mathfrak{G}^{\alpha\beta}\delta g_{\alpha\beta}
 - \frac{v^\mu}{n v_s}\Lambda T\delta s.
\end{equation*}
Notice that ${\bf X}$ vanishes when ${\bf W}$, $\delta s$ and $\delta{\bf  g}$ are zero, and in this case Eq.~(\ref{Eq:LinearizedFlow}) reduces to the homogeneous wave equation~(\ref{Eq:LinearizedPotentialFlow}) describing a linearized potential flow. In the general case, the one-form ${\bf W}$ and the variation of the entropy, $\delta s$, are obtained by integrating the advection equations~(\ref{Eq:WEvol},\ref{Eq:sEvol}), and therefore, they can be considered to be known quantities in the expression for the vector field ${\bf X}$. However, the variation of the metric fields, $\delta{\bf g}$, are coupled to the potential $\Psi$ through the lower order terms on the right-hand side of Eq.~(\ref{Eq:LinearizedFlow}) and through the linearized Einstein field equations.

%%%%%%%%%%%%%%%%%%%%%%%%%%%%%%%%%%%%%%%%%%%
\subsection{Fluid perturbations on a spherically symmetric background}
\label{SubSec:LinFluidSphBack}

So far, we have only assumed the background to be vorticity-free and isentropic. From now on, we assume in addition the background to be spherically symmetric. In particular, this implies that the four-velocity vector ${\bf u}$ has no angular components; hence it can be regarded as a vector field on the two-dimensional manifold $\tilde{M}$ and our zero vorticity assumption follows from the spherical symmetry. Defining \cite{cGjM00} the additional vector field ${\bf w}$ whose components are $w^a = -\tilde{\varepsilon}^a{}_b u^b$, we obtain an orthonormal basis of vector fields satisfying $\tilde{g}_{ab} = -u_a u_b + w_a w_b$ and $\tilde{\varepsilon}_{ab} = -2u_{[a} w_{b]}$. The connection coefficients with respect to this basis are entirely determined by the two quantities $\mu := \tilde{\nabla}^a u_a$ and $\nu := \tilde{\nabla}^a w_a$. 

Since the one-form ${\bf W}$ is orthogonal to ${\bf u}$, we can expand it according to
\begin{equation*}
W_a = U w_a,\qquad
W_A = \hat{\nabla}_A V + \hat{\varepsilon}_A{}^B\hat{\nabla}_B\omega,
\end{equation*}
where $U$, $V$, and $\omega$ are angular-dependent scalars on $\tilde{M}$, and where we have used the decomposition described in Eq.~(\ref {Eq:1FormDecomp}). Performing the calculations in the Regge-Wheeler gauge, for which $\delta g_{ab} = H_{ab}$, $\delta g_{aB} = \hat{\varepsilon}_B{}^C\hat{\nabla}_C h_a$, $\delta g_{AB} = r^2\hat{g}_{AB} J/2$, we first obtain from Eqs.~(\ref{Eq:JVar},\ref{Eq:nhVar}) and (\ref{Eq:hVar}),
\begin{eqnarray*}
\frac{\delta h}{h} &=& -\frac{1}{h}\pounds_{\bf u} \Psi + \frac{1}{2} H_{ab} u^a u^b,\\
\mathfrak{J}^a 
 &=& \frac{n}{h}\left[ \mathfrak{g}^{ab}\left( \tilde{\nabla}_b\Psi + U w_b - H_{bc} v^c \right)
  + \frac{1}{2}(\sigma^{cd} H_{cd} + J) v^a 
  - u^a\frac{\Lambda}{v_s^2} T\delta s \right],\\
 \mathfrak{J}^A &=& \frac{1}{r^2}\frac{n}{h}\left[ \hat{\nabla}^A(\Psi + V)
 + \hat{\varepsilon}^{AB}\hat{\nabla}_B (\omega - h_a v^a) \right],
\end{eqnarray*}
where we have defined the sonic two-metric
\begin{equation}
\mathfrak{g}^{ab} := \tilde{g}^{ab} + \left(1 - \frac{1}{v_s^2} \right) u^a u^b.
\label{Eq:SonicMetric2D}
\end{equation}
Therefore, the linearized continuity equation gives
\begin{equation}
-\tilde{\nabla}_a\left( r^2\frac{n}{h}\mathfrak{g}^{ab}\tilde{\nabla}_b\Psi \right)
 - \frac{n}{h}\hat{\Delta}\Psi
 = \tilde{\nabla}_a\left\{ r^2\frac{n}{h}\left[ U w^a - \mathfrak{g}^{ab} H_{bc} v^c
  + \frac{1}{2} v^a(\sigma^{bc} H_{bc} + J) 
  - u^a\frac{\Lambda}{v_s^2} T\delta s\right] \right\} 
  + \frac{n}{h}\hat{\Delta} V.
\label{Eq:LinearizedFlowSS}
\end{equation}
The right-hand side of this equation could be further simplified by using the background equation $\tilde{\nabla}_a(r^2 n u^a) = \tilde{\nabla}_a(r^2 n h^{-1} v^a) = 0$. For a spherically symmetric background, Eqs.~(\ref{Eq:WEvol}) and (\ref{Eq:sEvol}) imply the ordinary differential equation
\begin{equation}
\pounds_{\bf u} \omega = 0,
\label{Eq:LinEulerOdd}
\end{equation}
in the odd-parity sector, and in the even-parity sector the system of differential equations
\begin{equation}
(\pounds_{\bf u} + \mu) U = T(\delta s)',\qquad
\pounds_{\bf u} V = T\delta s,\qquad
\pounds_{\bf u}(\delta s) = 0,
\label{Eq:LinEulerEven}
\end{equation}
with $(\delta s)' = w^a\tilde{\nabla}_a(\delta s)$.

In the next three subsections we combine the above results with the equations describing gravitational perturbations derived in Sec.~\ref{Sec:Mat}, obtaining effective equations describing the linear fluctuations of a spherically symmetric self-gravitating fluid.

%%%%%%%%%%%%%%%%%%%%%%%%%%%%%%%%%%%%%%%%%%%%
\subsection{The linearized stress-energy tensor}
\label{SubSec:LinFluidSET}

The variation of the stress-energy tensor $T_{\mu\nu} = n h^{-1} v_\mu v_\nu + p g_{\mu\nu}$ yields the general expression
\begin{eqnarray*}
\delta T_{\mu \nu} &=&
n\left[ g_{\mu \nu} + \left( \frac{1}{v_s^2} - 1\right) u_{\mu} u_{\nu} \right]
\left[ \frac{1}{2}u^{\alpha}v^{\beta} \delta g_{\alpha \beta} - \pounds_{\bf u}\Psi  \right]\\
 &-& n\left[ g_{\mu\nu} + \frac{\Lambda}{v_s^2} u_{\mu} u_{\nu} \right] T\delta s 
  + 2n u_{(\mu}\left[ \nabla_{\nu)}\Psi + W_{\nu)} \right] + p\delta g_{\mu \nu},
\end{eqnarray*}
where we have used Eqs.~(\ref{Eq:nhVar},\ref{Eq:hVar},\ref{Eq:vDecomp}) and the first law of thermodynamics $dp = n dh - n T ds$. For a spherically symmetric background, the corresponding quantities $\tau_{ab}$, $\mu_a$, $\nu_a$, $\alpha$, $\lambda$ and $\beta$ defined in Eqs.~(\ref{Eq:LinTmunu}) are
\begin{equation}
\nu_a = p h_a + n u_a \omega,\qquad \beta = 0,
\label{Eq:LinTmunuOdd}
\end{equation}
in the odd parity sector, and
\begin{subequations}
\label{Eq:LinTmunuEven}
\begin{eqnarray}
\tau_{ab} &=&
n\left[ \tilde{g}_{ab} + \left( \frac{1}{v_s^2} - 1\right) u_a u_b \right]
\left[ \frac{1}{2}u^c v^d H_{cd} - u^c\tilde{\nabla}_c\Psi \right]
 - n\left[ \tilde{g}_{ab} + \frac{\Lambda}{v_s^2} u_a u_b \right] T\delta s 
\nonumber\\
 &+& 2n u_{(a}\left[ \tilde{\nabla}_{b)}\Psi + U w_{b)} \right] + p H_{ab},
\label{Eq:LinTmunuEven1}\\
\mu_a &=& n u_a(\Psi + V),
\label{Eq:LinTmunuEven2}\\
\alpha &=& 0,
\label{Eq:LinTmunuEven3}\\
\lambda &=& n\left[ u^a v^b H_{ab} - 2(u^a\tilde{\nabla}_a\Psi + T\delta s) \right]
 + p J,
\label{Eq:LinTmunuEven4}
\end{eqnarray}
\end{subequations}
in the even parity sector.

%%%%%%%%%%%%%%%%%%%%%%%%%%%%%%%%%%%%%%%%%%%%
\subsection{The linearized Einstein-Euler equations: odd-parity sector}
\label{SubSec:LinFluidOdd}

The odd-parity perturbations of a spherically symmetric fluid configuration are described by the linearized Euler equation~(\ref{Eq:LinEulerOdd}) and the linearized Einstein equations~(\ref{Eq:LinEinsteinOdd}), with the coefficients $\nu_a$ and $\beta$ given in Eq.~(\ref{Eq:LinTmunuOdd}), which yield
\begin{subequations}
\begin{eqnarray}
\tilde{*} d(r^2 {\cal F}) - (\hat{\Delta} + 2) {\bf h} &=& 2\kappa r^2 n\underline{\bf u}\omega,
\label{Eq:LinEinsteinOddFluid1}\\
\tilde{d}^\dagger {\bf h} &=& 0,
\label{Eq:LinEinsteinOddFluid2}
\end{eqnarray}
\end{subequations}
where ${\cal F} = r^2\tilde{*} d(r^{-2} {\bf h})$ and $\underline{\bf u} := u_a dx^a$ is the one-form on $\tilde{M}$ corresponding to the four-velocity of the background flow. We have already noticed in Sec.~\ref{Sec:Bianchi} that Eq.~(\ref{Eq:LinEinsteinOddFluid2}) follows from Eq.~(\ref{Eq:LinEinsteinOddFluid1}) and the divergence law for the stress-energy tensor. In the present case, this can also be verified explicitly by applying the co-differential on both sides of Eq.~(\ref{Eq:LinEinsteinOddFluid1}), using $\pounds_{\bf u}\omega = 0$ and the background equation $\tilde{d}^\dagger(r^2 n\underline{\bf u}) = 0$.

A wave equation for the scalar field $\Phi := r{\cal F}$ follows from the method described in Sec.~\ref{SubSec:OddParityMatter}. For this, we use the background equation $d{\bf v} = 0$ and first note that
\begin{equation*}
\tilde{*} d{\bf \Upsilon} = 
\tilde{*} d (n\underline{\bf u}\omega) = \tilde{*} d\left( \frac{n}{h}\omega{\bf v}\right)
 = \tilde{*}\left[ h\left( \frac{n}{h}\omega \right)'
  \underline{\bf w}\wedge\underline{\bf u} \right] 
 = -h\left( \frac{n}{h}\omega \right)',
\end{equation*}
where the prime refers to the directional derivative along ${\bf w}$. With this observation, Eq.~(\ref{Eq:LinEinsteinOddMatterF}) yields (cf. Appendix B in Ref.~\cite{kTaC67}):
\begin{equation*}
r^4\tilde{d}^\dagger\left[ \frac{1}{r^2} d(r^2{\cal F}) \right] 
 - (\hat{\Delta} + 2)(r^2{\cal F}) 
 = -2\kappa r^4 h\left( \frac{n}{h}\omega \right)'.
\end{equation*}
Introducing ${\cal F} = \Phi/r$ and using the background equation~(\ref{Eq:EinsteinSph2}), we can rewrite this equation in the form
\begin{equation}
\tilde{\Box}\Phi + \left[ -\frac{\hat{\Delta}}{r^2} - \frac{6m}{r^3} 
 + \frac{\kappa}{2}(\epsilon - p) \right]\Phi 
 = -2\kappa r h\left( \frac{n}{h}\omega \right)',
\label{Eq:ThorneCampolattaroBis}
\end{equation}
which generalizes the Regge-Wheeler equation~(\ref{Eq:ReggeWheeler}) to the case of odd-parity linear fluctuations of self-gravitating spherical fluid configurations. Once $\Phi$ is known, the metric perturbation ${\bf h}$ can be reconstructed from Eq.~(\ref{Eq:LinEinsteinOddFluid1}) on the mono-dipole-free space.

%%%%%%%%%%%%%%%%%%%%%%%%%%%%%%%%%%%%%%%%%%%%
\subsection{The linearized Einstein-Euler equations: even-parity sector}
\label{SubSec:LinFluidEven}

The even-parity perturbations are described by the linearized fluid equations~(\ref{Eq:LinearizedFlowSS},\ref{Eq:LinEulerEven}) and the linearized Einstein field equations~(\ref{Eq:LinEinsteinEven}), where the coefficients $\tau_{ab}$ and $\mu_a$, $\alpha$ and $\lambda$ are given in Eqs.~(\ref{Eq:LinTmunuEven}) above. First of all, we recall that the Eqs.~(\ref{Eq:LinEulerEven}) are simple advection equations along the flow lines, which are decoupled from the remaining perturbation equations. Therefore, specifying initial data for $(U,V,\delta s)$ on an initial hypersurface $\Sigma_0$ completely determines the evolution of these quantities. Next, as discussed in Sec.~\ref{SubSec:Evolution}, the linearized Einstein equations give rise to a system of two wave equations for the metric fields $(C,D)$, see Eqs.~(\ref{Eq:CDEvol1+1}), where the source terms $\Gamma$ and $\Delta$ are determined by the linearized stress-energy tensor. Here, the vector fields ${\bf u}$ and ${\bf w}$ which were introduced in Sec.~\ref{SubSec:Evolution} to determine the null vectors ${\bf k} := {\bf u} + {\bf w}$ and ${\bf l} := {\bf u} - {\bf w}$ are oriented such that ${\bf u}$ coincides with the four-velocity of the background fluid flow. Using the definitions in Eqs.~(\ref{Eq:DefGammaDelta}) and the expressions~(\ref{Eq:LinTmunuEven}), we find
\begin{eqnarray*}
\Gamma &=& 
\frac{n}{2}\left( \frac{1}{v_s^2} - 1\right) \left[ \frac{h}{2}(C+D) - \dot{\Psi} \right] 
 - \frac{n}{2}\frac{\Lambda}{v_s^2} T\delta s + nh C
 + \left[ \frac{1}{r^2}\tilde{\nabla}_{\bf l} \left( r^2n \right) - (\mu - \nu)n \right] (\Psi + V)
 + n \left( \tilde{\nabla}_{\bf l} V + U \right),\\
\Delta &=&
 \frac{n}{2}\left( \frac{1}{v_s^2} - 1\right) \left[ \frac{h}{2}(C+D) - \dot{\Psi} \right] 
 - \frac{n}{2}\frac{\Lambda}{v_s^2} T\delta s + nh D 
 + \left[ \frac{1}{r^2}\tilde{\nabla}_{\bf k} \left( r^2n \right) - (\mu + \nu)n \right] (\Psi + V)
+ n \left( \tilde{\nabla}_{\bf k} V - U \right),
\end{eqnarray*}
where $\Lambda$ is defined in Eq.~(\ref{Eq:LambdaDef}). Notice that Eq.~(\ref{Eq:CEvol1+1}) is transformed into Eq.~(\ref{Eq:DEvol1+1}) one under the symmetry~(\ref{Eq:NestorSym}) augmented by $U\mapsto -U$.

Together with the evolution equation for $\Psi$, Eq.~(\ref{Eq:LinearizedFlowSS}), we obtain a closed wave system for the three quantities $(C,D,\Psi)$. In order to write down the evolution equation for $\Psi$, we notice that $J$ can be eliminated from the right-hand side of Eq.~(\ref{Eq:LinearizedFlowSS}) by virtue of Eq.~(\ref{Eq:LinEinsteinEven2}) and the satisfaction of the background equation $\tilde{\nabla}_a(r^2 n u^a) = 0$. This yields the equation
\begin{eqnarray}
 -\tilde{\nabla}_a\left(\mathfrak{g}^{ab}\tilde{\nabla}_b\Psi \right)
&-& \left[ \tilde{\nabla}_a\log\left( \frac{r^2 n}{h} \right) \right]
\mathfrak{g}^{ab}\tilde{\nabla}_b\Psi
 - \frac{\hat{\Delta}}{r^2}(\Psi + V)
 - 2\kappa n h(\Psi + V)
\nonumber\\
 &=& \frac{h}{2}u^a \tilde\nabla_a
\biggl[\left(\frac{1}{v_s^2} - 1 \right)( C+D) - \frac{2\Lambda}{h v_s^2} T\delta s\biggr]
\nonumber\\
 &+& h\left[ \frac{\tilde{\nabla}_{\bf k}(r^2 n)}{r^2 n} - (\mu + \nu) \right] C
  + h\left[ \frac{\tilde{\nabla}_{\bf l}(r^2 n)}{r^2 n} - (\mu - \nu) \right] D
  + U' + \nu U + \left[\log\left( \frac{r^2 n}{h} \right) \right]' U,
\label{Eq:LinearizedFlowSCD}
\end{eqnarray}
describing the evolution of the fluid field $\Psi$. Here, $\mathfrak{g}^{ab}$ refers to the two-dimensional sonic metric on $\tilde{M}$, see Eq.~(\ref{Eq:SonicMetric2D}).

The evolution system~(\ref{Eq:CDEvol1+1},\ref{Eq:LinearizedFlowSCD}) for $(C,D,\Psi)$ is subject to the restriction in Eq.~(\ref{Eq:Rel_CD1}), where the quantity $G$ is
\begin{equation*}
G = [n(\Psi + V)]' + n\nu(\Psi + V),
\end{equation*}
for the fluid case. Notice that $G$ does not depend on $J$; hence, Eq.~(\ref{Eq:Rel_CD1}) gives a relation between $(C,D)$ and the matter fields $(\Psi,V)$. The quantity $J$ can be reconstructed from the variables $(C,D,\Psi,U,V,\delta s)$ using Eq.~(\ref{Eq:JEliminate}) where
\begin{equation*}
F = -\frac{n}{2}\left( \frac{1}{v_s^2} + 1\right) \left[ \frac{h}{2}(C+D) - \dot{\Psi} \right]
 - \frac{n}{2}\left( 2 - \frac{\Lambda}{v_s^2} \right) T\delta s
  + \left[ \frac{1}{r^4}\tilde{\nabla}_{\bf u}(r^4 n) + \mu n \right] (\Psi + V)+ n\dot{V}.
\end{equation*}

Initial data for the evolution system~(\ref{Eq:CDEvol1+1},\ref{Eq:LinearizedFlowSCD})  can be constructed by specifying $(J,\dot{J},\Psi,\dot{\Psi},U,V,\delta s)$ on an initial hypersurface $\Sigma_0$, and by following the algorithm described in Sec.~\ref{SubSec:InitialData}. Namely, we first set
\begin{eqnarray*}
a_{-2}^+ C &:=& -\frac{1}{4}a_0^- J  + \kappa n(\Psi + V),\\
a_2^- D &:=& -\frac{1}{4}a_0^+ J + \kappa n(\Psi + V).
\end{eqnarray*}
Next, we solve the ordinary differential equations~(\ref{Eq:ODECD}) for $(C,D)$, where the matter source terms are given by the expressions $\Gamma - F - G = X - Y + nh C$, $\Delta - F + G = X + Y + nh D$ with
\begin{equation*}
X := \frac{n}{v_s^2}\left[ \frac{h}{2}(C+D) - \dot{\Psi} \right] 
 + n\left( 1 - \frac{\Lambda}{v_s^2} \right) T\delta s
 - 2n\left( \frac{\dot{r}}{r} +Ê\mu \right)(\Psi + V),\qquad
Y:= n(\Psi' + 2V' - U) + \frac{2}{r}(rn)'(\Psi + V).
\end{equation*}
Finally, $(\dot{C},\dot{D})$ are determined by Eq.~(\ref{Eq:CdotDdot}). This provides initial data for the wave system~(\ref{Eq:CDEvol1+1},\ref{Eq:LinearizedFlowSCD}). At this point, we recall the freedom described in Eq.~(\ref{Eq:FluidGT}) that allows to set the initial value for $\Psi$ to zero. Therefore, our initial data contains six degrees of freedom, in accordance with the results in~\cite{cGjM00}.

%%%%%%%%%%%%%%%%%%%%%%%%%%%%%%%%%%%%%%%%%%%%%
\section{Summary and conclusions}
\label{Sec:Conclusions}
%%%%%%%%%%%%%%%%%%%%%%%%%%%%%%%%%%%%%%%%%%%%%

In this paper we have presented a gauge-invariant perturbation formalism for spherically symmetric background configurations in general relativity, without assuming the background to be static or vacuum. The new feature of this formalism is that it combines the covariant, gauge-invariant method in~\cite{uGuS79} with the quasilocal approach in~\cite{jJ99}, so that it does not require a decomposition of the perturbations into tensor spherical harmonics. 

As a pedagogical first step, we have derived effective wave equations on the two-dimensional orbit manifold \hbox{$\tilde{M} = M\slash SO(3)$} orthogonal to the two-spheres, describing the propagation of scalar and electromagnetic test fields on an arbitrary spherically symmetric background. Then, we have further developed our formalism to describe linear metric fluctuations. Our approach is based on the $2+2$ form of the background geometry and the construction of a full set of gauge-invariant amplitudes. These amplitudes are angular-dependent tensor fields on $\tilde{M}$, which behave as scalars under rotations of the metric two-spheres. We have expressed the linearized Einstein equations in terms of these gauge-invariant quantities which decouple into two groups, describing perturbations with {\em odd} and {\em even} parity, respectively.

Next, we have applied our formalism to the vacuum case and have derived the covariant generalizations of the Regge-Wheeler and Zerilli master equations, which describe the propagation of arbitrary linearized gravitational waves on a Schwarzschild black hole. In particular, we have shown that the Regge-Wheeler equation can easily be obtained in both, the odd- {\em and the even-}parity sectors, although in the even-parity sector it does not represent a master equation in the strict sense, since in this case an additional differential equation needs to be solved in order to reconstruct the metric perturbations.

For a Schwarzschild background, the master equations describing the propagation of scalar, electromagnetic, and linearized gravitational perturbations have the form of an effective wave equation on $(\tilde{M},\tilde{\bf g})$,
\begin{equation}
\tilde{\Box}\Phi + V(r)\Phi = {\cal S},\qquad
V(r) := \frac{1}{r^2}\left( -\hat{\Delta} + B(r) \right),
\label{Eq:MasterEquation}
\end{equation}
where $\tilde{\Box}$ is the covariant d'Alambertian on the radial part $(\tilde{M},\tilde{\bf g})$ of the Schwarzschild-Kruskal manifold, $\hat{\Delta}$ denotes the Laplacian on the sphere $S^2$, $r$ is the areal radius, and the operator $B(r)$ is defined as
\begin{displaymath}
B(r) := \left\{ \begin{array}{ll} 
 \frac{2m}{r} + \mu^2 r^2, & \hbox{for a Klein-Gordon field of mass $\mu$}\\
 0, & \hbox{for odd- and even-parity electromagnetic fields}\\
-\frac{6m}{r}, & \hbox{for odd-parity gravitational perturbations}\\
 -\frac{6m}{r} - \frac{24m}{r}\left( 1 - \frac{3m}{r} \right) A(r)^{-1} 
 + \frac{72m^2}{r^2}\left( 1 - \frac{2m}{r} \right) A(r)^{-2},
 & \hbox{for even-parity gravitational perturbations}
 \end{array} \right.
\end{displaymath}
with $A(r) := -\hat{\Delta} - 2 + 6m/r$. Here, the source term ${\cal S}$ is zero for the Klein-Gordon field, and in the electromagnetic case it is determined by the four-current charge density, see Eqs.~(\ref{Eq:MaxwellOdd},\ref{Eq:MaxwellMasterEq}). In the gravitational case, ${\cal S}$ is zero for vacuum perturbations. For gravitational perturbations generated by an infinitesimal stress-energy tensor, the expressions for ${\cal S}$ are given in~\cite{kMeP05}. Recall that in the electromagnetic and gravitational cases, $\Phi$ is restricted to the monopole-free and the mono-dipole-free spaces, respectively. This means that the expansion of $\Phi$ in spherical harmonics only contains terms with angular momentum number $\ell\geq L$ with $L=0$ in the scalar, $L=1$ in the electromagnetic and $L=2$ in the gravitational cases. Based on the energy-type estimates described in~\cite{mDiR08}, one can prove that the solutions of Eq.~(\ref{Eq:MasterEquation}) belonging to sufficiently regular initial data on a spacelike slice remain uniformly bounded outside the black hole.

Next, we considered metric perturbations coupled to arbitrary matter fields. While in the odd-parity sector the generalization to matter is rather straightforward in most cases, in the even-parity sector it is unclear whether or not the Zerilli approach works. Recall that in vacuum, Zerilli's method in the covariant approach is based on the observation that one of the linearized Einstein equations implies the zero exterior derivative of the Zerilli one-form ${\bf Z}$, see Eq.~(\ref{Eq:LinEvenVac1}), which means that ${\bf Z}$ can be written as the differential of a scalar potential. However, with the exception of particular matter models, obtaining a closed one-form does not seem possible in the general case. Therefore, we proposed an alternative approach which naturally leads to a constrained wave system for two gauge-invariant metric perturbation amplitudes. In vacuum, this wave system decouples into two wave-like equations which are related to the Teukolsky equations~\cite{sT72,jBwP73} for the Weyl scalars $\Psi_2$ and $\Psi_{-2}$. In the presence of matter fields, the two equations are coupled to each other through the linearized stress-energy tensor, and, together with suitable evolution equations for the matter fields, describe the propagation of the linearized gravitational and matter field perturbations.

As an application, we considered linear perturbations of a spherically symmetric perfect fluid configuration. Assuming local thermodynamic equilibrium, and assuming that the background is isentropic, we decomposed the fluid perturbations into two parts. One part describes a linearized potential flow and is entirely determined by a scalar quantity $\Psi$. The complementary part is described by a vector field ${\bf W}$ which is orthogonal to the four-velocity of the background flow, and generates a non-trivial vorticity field at the linearized level. The evolution of ${\bf W}$ couples only to the evolution of the perturbed entropy, and the corresponding equations are simple advection equations along the background flow. Then, we combined these equations with those describing gravitational perturbations and obtained effective equations on $\tilde{M}$ describing linear fluctuations of a spherically symmetric self-gravitating fluid. In the odd-parity sector we obtained an inhomogeneous master equation generalizing the Regge-Wheeler equation to the fluid case where the source term is determined by the fluid field ${\bf W}$. In the even-parity sector, we obtained a constrained, coupled wave system for two gauge-invariant metric perturbation amplitudes and the fluid potential $\Psi$. This evolution system may be solved using numerical methods, regarded as a Cauchy problem with initial data constructed according to the algorithm given in Sec.~\ref{SubSec:InitialData}. 

The covariant propagation equations found in this article should have many applications in astrophysics. One example is a detailed investigation of radial accretion flows (for models on fixed backgrounds see, for instance, Refs.~\cite{fM72,eCoS12}) concerning their stability and their quasi-normal oscillations. Another important application is the stability analysis of Cauchy horizons found in some spherically symmetric models, like a Reissner-Nordstr\"om black hole or a spherical dust cloud undergoing complete gravitational collapse. The latter is particularly interesting, since it it known to lead to the formation of globally naked shell-focusing singularities which are stable with respect to spherical perturbations~\cite{dC84,nOoS11}.

The fact that our formalism eliminates the need of decomposing the perturbations into spherical tensor harmonics represents a simplification with respect to previous approaches, which could be useful for a generalization to second- or higher-order perturbations. For example, the equations governing the second-order perturbations contain source terms which depend quadratically on the first-order ones, and when expanded into spherical harmonics this leads to the computation of Clebsch-Gordan coefficients. In our approach, this computation is unnecessary. Instead, the source terms need to be decomposed according to Eqs.~(\ref{Eq:VectorDecomp},\ref{Eq:TensorDecomp}) which can be performed by solving the elliptic equations~(\ref{Eq:Elliptic_fg},\ref{Eq:Elliptic_hk}) on the two-spheres.

Another potential advantage of the results derived in this article is related to the constrained wave system describing the gravitational perturbations in the even-parity sector. Since one of the scalars involved in this system is closely related to the Weyl scalar $\Psi_{-2}$, describing the outgoing gravitational radiation at future null infinity, the computation of the gravitational waves emitted by an isolated system consisting of a known spherically symmetric spacetime plus a linear perturbation should be rather direct. Potentially, this system could also be used as a wave extraction algorithm in the far field of a more complicated isolated system, like the coalescence of black holes or neutron stars. However, a persistent challenge in this case is an adequate identification of the spherically symmetric background. In fact, the gauge-invariant formalism is based on the knowledge of the background {\em including} its $2+2$ splitting $M = \tilde{M}\times S^2$, which requires the correct identification of the invariant two-spheres $S^2$.

Finally, it should also be possible to generalize our formalism to higher-dimensional, $SO(q+1)$-symmetric spacetimes of the form $M = \tilde{M}\times S^q$, where the dimensions of $\tilde{M}$ and the invariant spheres, $q=2,3,\ldots$, are arbitrary. As long as $q=2$, we can base the construction of gauge-invariant tensor fields on $\tilde{M}$ on the same decomposition as in Eqs.~(\ref{Eq:VectorDecomp},\ref{Eq:TensorDecomp}), although the effective equations obtained on $\tilde{M}$ are now more complicated when $\tilde{M}$ has dimension greater than two. When $q > 2$, one has to consider the decompositions~(\ref{Eq:VectorDecompBis},\ref{Eq:TensorDecompBis}) instead, and in this case, the equations decouple into scalar, vector, and tensor perturbations. See Refs.~\cite{kHiA04,aIhK11} for applications to the stability of higher-dimensional static black holes.

%%%%%%%%%%%%%%%%%%%%%%%%%%%
%%%   ACKNOWLEDGMENTS
%%%%%%%%%%%%%%%%%%%%%%%%%%%

\acknowledgments

It is a pleasure to thank T. Zannias for insightful discussions and comments and D. Brizuela for reading the manuscript. This work was supported in part by Grants CIC 4.19 to Universidad Michoacana and CONACyT Grants No. 46521, 101353 and 238758.

\appendix
%%%%%%%%%%%%%%%%%%%%%%%%%%%%%%%%%%%%%%%%%%%%%
\section{Covariant, orthonormal decomposition of one-forms and trace-free symmetric tensor fields on compact two-surfaces}
\label{App:Decomposition}
%%%%%%%%%%%%%%%%%%%%%%%%%%%%%%%%%%%%%%%%%%%%%

Let $(S,{\bm g})$ be a compact, oriented, two-dimensional Riemannian manifold without boundary which has positive Gauss curvature (By the Gauss-Bonnet theorem, this implies that $S$ is topologically equivalent to the two-sphere $S^2$.) Let ${\bm\varepsilon}$ be the associated volume form on $(S,{\bm g})$ and introduce the following scalar products:
\begin{equation}
< f,g > := \int\limits_S f\cdot g\, {\bm\varepsilon},\qquad
<{\bm\omega},{\bm\eta} > 
 := \int\limits_S g^{AB} \omega_A\cdot\eta_B\, {\bm\varepsilon},\qquad
<{\bm X},{\bm Y} > 
 := \int\limits_S g^{AB} g^{CD} X_{AC}\cdot Y_{BD}\, {\bm\varepsilon},
\label{Eq:DefScalarProd}
\end{equation}
for smooth functions $f$, $g$, smooth one-forms ${\bm\omega}$, ${\bm\eta}$ and smooth tensor fields ${\bm X}$, ${\bm Y}$ on $S$. In this appendix we show:

\begin{proposition}
\label{Prop:Decomposition}
Let ${\bm\omega}$ and ${\bm X}$ denote a $C^\infty$ one-form and symmetric, trace-less tensor field, respectively, on $(S,{\bm g})$. Then, there exist $C^\infty$ functions $f$, $g$, $h$, $k$ on $S$ such that
\begin{eqnarray}
&& \omega_A = \nabla_A f + \varepsilon_A{}^B\nabla_B g,
\label{Eq:VectorDecomp}\\
&& X_{AB} = (\nabla_A\nabla_B h)^{TF} + \varepsilon_{(A}{}^C\nabla_{B)}\nabla_C k,
\label{Eq:TensorDecomp}
\end{eqnarray}
where $Y_{AB}^{TF} := Y_{AB} - \frac{1}{2} g_{AB} g^{CD} Y_{CD}$ refers to the trace-free part of the tensor field ${\bm Y}$ on $S$.

Furthermore, the decompositions (\ref{Eq:VectorDecomp},\ref{Eq:TensorDecomp}) are unique and orthogonal with respect to the scalar products defined in (\ref{Eq:DefScalarProd}). The functions $f$ and $g$ are unique up to an additive constant and the constants $h$ and $k$ unique up to the addition of an element in the kernel of the operator $(\nabla_A\nabla_B)^{TF}$.
\end{proposition}

{\bf Remarks}:
\begin{enumerate}
\item The functions $(f,g)$ in Eq.~(\ref{Eq:VectorDecomp}) can be found by solving the elliptic equations
\begin{equation}
\Delta f = \nabla^A\omega_A,\qquad
\Delta g = -\varepsilon^{AB}\nabla_A\omega_B,
\label{Eq:Elliptic_fg}
\end{equation}
and the functions $(h,k)$ in Eq.~(\ref{Eq:TensorDecomp}) by solving
\begin{equation}
\Delta(\Delta + 2\hat{k}) h = 2\nabla^A \nabla^B X_{AB},\qquad
\Delta(\Delta + 2\hat{k}) k = -2\nabla^A \nabla^B \varepsilon_A{}^C X_{BC},
\label{Eq:Elliptic_hk}
\end{equation}
with $\Delta = \nabla^A\nabla_A$ the Laplacian and $\hat{k}$ the Gauss curvature with respect to $(M,{\bm g})$.
\item For the case of the unit two-sphere, $S = S^2$, the kernel of $(\nabla_A\nabla_B)^{TF}$ consists of the functions of the form $f = f_{00} Y^{00} + f_{11} Y^{10} + f_{11} Y^{11} + \bar{f}_{11} Y^{1-1}$ with $f_{00},f_{10}\in\Real$, $f_{11}\in\Complex$, and $Y^{\ell m}$ denoting the standard spherical harmonics. In the context of the perturbation theory described in this article, we call the constant part $f_{00} Y^{00}$ of $f = \sum f_{\ell m} Y^{\ell m}$ the {\em monopole} part, the part of $f$ which is equal to $f_{10} Y^{10} + f_{11} Y^{11} + \bar{f}_{11} Y^{1-1}$ the {\em dipole} part and the remaining part $f = \sum_{\ell\geq 2} f_{\ell m} Y^{\ell m}$ the {\em mono-dipole-free} part. In this case, it is not difficult to prove the proposition by decomposing the elliptic equations~(\ref{Eq:Elliptic_fg}) and (\ref{Eq:Elliptic_hk}) into spherical harmonics.
\item The decomposition~(\ref{Eq:VectorDecomp}) also follows from the Hodge decomposition theorem (see, for instance Proposition V.8.2 in Ref.~\cite{Taylor-Book}) and the fact that all harmonic one-form vanish on $(S,{\bm g})$ with our assumptions, see Lemma~\ref{Lem:IntegralIdentities} below.
\end{enumerate}

\proofof{proposition~\ref{Prop:Decomposition}} We split the proof in two parts. In the first part we show there exist $C^\infty$ functions $f$ and $h$ and a divergence-free $C^\infty$ one-forms ${\bm\eta}$ and a $C^\infty$ symmetric, trace-free tensor field ${\bm Y}$ satisfying $\nabla^A\nabla^B Y_{AB} = 0$ such that
\begin{eqnarray}
&& \omega_A = \nabla_A f + \eta_A,
\label{Eq:VectorDecompBis}\\
&& X_{AB} = (\nabla_A\nabla_B h)^{TF} + Y_{AB}.
\label{Eq:TensorDecompBis}
\end{eqnarray}
These decompositions are orthogonal with respect to the scalar products defined in Eq.~(\ref{Eq:DefScalarProd}); hence, they are unique (if they exist). The first equation implies that $f$ must be such that the one-form ${\bm\omega} - \nabla f$ is divergence-free, which is equivalent to the statement 
\begin{equation}
< {\bm\omega} - \nabla f,\nabla u > = 0
\label{Eq:WeakProblem1}
\end{equation}
for all $C^\infty$ functions $u$ on $S$. Similarly, the function $h$ must satisfy the problem
\begin{equation}
< {\bm X} - (\nabla\nabla h)^{TF},\nabla\nabla v > = 0
\label{Eq:WeakProblem2}
\end{equation}
for all $C^\infty$ functions $v$ on $S$. In order to show that the problems~(\ref{Eq:WeakProblem1},\ref{Eq:WeakProblem2}) possess a solution, we use standard tools from the theory of Fredholm operators, see, for instance, Ref.~\cite{Kato-Book}, or Theorem 4 in appendix A of~\cite{gNoS06} for a summary of the relevant results.

Denote by $H^m(S)$ the Sobolev space consisting of the set of square-integrable functions on $S$ with square-integrable weak derivatives of order smaller than or equal to $m$ (see, for instance, Ref.~\cite{Taylor-Book} for more details). We define for each $\lambda\geq 0$ the following bounded, symmetric quadratic forms,
\begin{eqnarray*}
&& P_\lambda : H^1(S) \times H^1(S) \to \Real, \quad
      (f,u)\mapsto P_\lambda(f,u) := < \nabla f,\nabla u > + \lambda < f,u >,\\
&& Q_\lambda : H^2(S) \times H^2(S) \to \Real, \quad
      (h,v)\mapsto Q_\lambda(h,v) := < (\nabla\nabla h)^{TF}, (\nabla\nabla v)^{TF} >
       + \lambda < \nabla h,\nabla v > + \lambda < h,v >,
\end{eqnarray*}
and the bounded linear functionals
\begin{eqnarray*}
&& J_{\bm\omega} : H^1(S) \to \Real,\quad 
      u\mapsto J_{\bm\omega}(u) := < {\bm\omega}, \nabla u >,\\
&& K_{\bm X} : H^2(S) \to \Real,\quad 
      v\mapsto K_{\bm X}(u) := < {\bm X}, (\nabla\nabla v)^{TF} >,
\end{eqnarray*}
With these notations, we reformulate the problems described in Eqs.~(\ref{Eq:WeakProblem1},\ref{Eq:WeakProblem2}) as finding $f\in H^1(S)$ and $h\in H^2(S)$ such that
\begin{eqnarray*}
P_0(f,u) &=& J_{\bm\omega}(u) \hbox{ for all $u\in H^1(S)$},\\
Q_0(h,v) &=& K_{\bm X}(v) \hbox{ for all $v\in H^2(S)$}.
\end{eqnarray*}

Notice that for $\lambda = 1$, $P_1$ is the standard scalar product on $H^1(S)$. Therefore, it follows by the Riesz representation lemma that the map $L_\lambda : H^1(S) \to (H^1(S))^*$, $f\mapsto P_\lambda(f,\cdot)$ is invertible for $\lambda = 1$. In particular, it is a Fredholm operator of index zero. On the other hand, by compact embedding, the difference operator $L_1 - L_0$ is compact, which implies that also the map $L_0$ is Fredholm with index zero. Therefore, the equation $L_0 f = J_{\bm\omega}$ has a solution if and only if $J_{\bm\omega}(u)=0$ for all $u\in H^1(S)$ such that $\nabla u = 0$. Since this is always the case by the definition of $J_{\bm\omega}$, the existence of a solution $f\in H^1(S)$ satisfying Eq.~(\ref{Eq:WeakProblem1}) follows. By elliptic regularity theory (see, for instance Ref.~\cite{Taylor-Book}) $f$ is $C^\infty$ smooth and $\eta_A := \omega_A - \nabla_A f$ is divergence-free.

Similarly, $Q_\lambda$ is a scalar product on $H^2(S)$ for sufficiently large $\lambda > 0$. In order to see this, we apply the integral identity~(\ref{Eq:IntIdentity1}) below to $\beta_A = \nabla_A h$, obtaining
\begin{displaymath}
Q_\lambda(h,h) 
 = \frac{1}{2}\int\limits_S (\nabla^A\nabla^B h)(\nabla_A\nabla_B h) {\bm\varepsilon}
 + \int\limits_S \left( \lambda - \frac{\hat{k}}{2} \right)(\nabla^A h)(\nabla_A h) {\bm\varepsilon}
 + \lambda\int\limits_S h^2\, {\bm\varepsilon},
\end{displaymath}
which shows positivity for $\lambda > \hat{k}/2$ and also proves that in this case the norm induced by $Q_\lambda$ is equivalent to the standard norm on $H^2(S)$. Therefore, for $\lambda > \hat{k}/2$ the maps $M_\lambda : H^2(S) \to (H^2(S))^*$, $h\mapsto Q_\lambda(h,\cdot)$ are invertible and Fredholm of index zero. Since $M_\lambda - M_0$ is compact, it follows that also $M_0$ is Fredholm with index zero. Therefore, the equation $M_0 h = K_{\bm X}$ has a solution if and only if $K_{\bm X}(v)=0$ for all  $v\in H^2(S)$ such that $(\nabla\nabla v)^{TF}=0$, which is always the case by the definition of $K_{\bm X}$. The existence of a solution $h\in H^2(S)$ satisfying Eq.~(\ref{Eq:WeakProblem2}) follows, and by elliptic regularity it is $C^\infty$ smooth. Then, $Y_{AB} := X_{AB} - (\nabla_A\nabla_B h)^{TF}$ satisfies the required property $\nabla^A\nabla^B Y_{AB} = 0$.

This concludes the first part of the proof and shows existence and uniqueness of the decompositions~(\ref{Eq:VectorDecompBis},\ref{Eq:TensorDecompBis}). Notice that so far, no topological restrictions on $S$ have been used.

In the second part of the proof we show the existence of $C^\infty$ functions $g$ and $k$ on $S$ satisfying $\eta_A = \varepsilon_A{}^B\nabla_B g$ and $Y_{AB} = \varepsilon_{(A}{}^C\nabla_{B)}\nabla_C k$. In order to prove this, we apply the decomposition~(\ref{Eq:VectorDecompBis}) to the one-form $\alpha_A := -\varepsilon_A{}^B\eta_A$. Therefore, $\alpha_A = \nabla_A g + \beta_A$ with $\nabla^A\beta_A = 0$. Since $0 = \nabla^A\eta_A = \nabla^A \varepsilon_A{}^B(\nabla_B g + \beta_B) = \varepsilon^{AB}\nabla_A\beta_B$, the one-form $\beta_B$ on $S$ is both divergence- and curl-free. By the integral identity~(\ref{Eq:IntIdentity1}) below, this implies ${\bm\beta} = 0$ since $S$ has positive Gauss curvature, and it follows that $\eta_A = \varepsilon_A{}^B\alpha_B = \varepsilon_A{}^B\nabla_B g$. By a similar argument, we use the decomposition~(\ref{Eq:TensorDecompBis}) and write $Y_{AB} = \varepsilon_{(A}{}^C(\nabla_{B)}\nabla_C k + Z_{B)C})$ where the $C^\infty$ symmetric, trace-free tensor field ${\bm Z}$ satisfies $\nabla^A\nabla^B Z_{AB} = 0$ and $0 = \nabla^A\nabla^B Y_{AB} = \nabla^A\nabla^B\varepsilon_A{}^C Z_{BC}$. Therefore, the one-form $\eta_A:=\nabla^B Z_{AB}$ is both divergence- and curl-free which implies ${\bm\eta} = 0$. According to the second integral identity~(\ref{Eq:IntIdentity2}) below, this implies ${\bm Z}=0$ which concludes the second part of the proof.
\qed

\begin{lemma}
\label{Lem:IntegralIdentities}
Let ${\bm\beta}$ and ${\bf Z}$ denote a $C^\infty$ one-form and symmetric, trace-free tensor field, respectively, on $(S,{\bf g})$. Then, the following integral identities hold:
\begin{eqnarray}
&& \int\limits_S \left[ (\nabla^A\beta^B)(\nabla_A\beta_B)
 + \hat{k} \beta^A\beta_A \right] {\bm\varepsilon}
 = \int\limits_S \left[ 
 \frac{1}{2}(\nabla^A\beta^B - \nabla^B\beta^A)(\nabla_A\beta_B - \nabla_B\beta_A) 
 + (\nabla^A\beta_A)^2 \right] {\bm\varepsilon},
\label{Eq:IntIdentity1}\\
&& \int\limits_S \left[ (\nabla^A Z^{BC})(\nabla_A Z_{BC}) 
 + 2\hat{k} Z^{AB} Z_{AB} \right] {\bm\varepsilon}
 = 2\int\limits_S g^{AB}(\nabla^C Z_{CA})(\nabla^D Z_{DB}) {\bm\varepsilon},
\label{Eq:IntIdentity2}
\end{eqnarray}
where $\hat{k}$ denotes the Gauss curvature of $(S,{\bf g})$. In particular if $\hat{k} > 0$ it follows that there are no non-trivial divergence- and curl-free one-forms on $(S,{\bf g})$ nor are there non-trivial symmetric, trace- and divergence-free tensor fields on $(S,{\bf g})$. 
\end{lemma}

\proof For the first identity, we integrate
\begin{equation*}
(\nabla^A\beta^B)(\nabla_A\beta_B) 
 = (\nabla^A\beta^B - \nabla^B\beta^A)(\nabla_A\beta_B) 
 + (\nabla^B\beta^A)(\nabla_A\beta_B) 
\end{equation*}
over $S$ and use Gauss' theorem twice to bring the second term on the right-hand side in the form of the square of the divergence of ${\bm\beta}$. For the second identity, we integrate
\begin{equation*}
(\nabla^A Z^{BC})(\nabla_A Z_{BC}) = (\varepsilon_{AB}\nabla^A Z^{BC})(\varepsilon^{EF}\nabla_E Z_{FC}) + (\nabla^B Z^{AC})(\nabla_A Z_{BC})
\end{equation*}
over $S$, use Gauss' theorem twice to rewrite the second term on the right-hand side
as the square of the divergence of ${\bm Z}$ and use the fact that $\varepsilon_A{}^C Z_{CB} = -Z_{AD}\varepsilon^D{}_B$.
\qed

\bibliographystyle{unsrt}
\bibliography{refs_fluid}
%%%%%%%%%%%%%%%%%%%%%%%%%%%%%%%%%%%%%%%%%%%%%

\end{document}